\documentclass[twocolumn]{aastex701}

\usepackage{amsmath}
\usepackage{amssymb}	
\usepackage{natbib}
\usepackage[figuresright]{rotating}
\usepackage{threeparttable}
\usepackage{subfigure}
\usepackage{soul}
\usepackage[normalem]{ulem}
\usepackage{cancel}

\hypersetup{colorlinks,linkcolor={blue},citecolor={blue},urlcolor={blue}}

\newcommand{\cmsq}{{cm$^{-2}$}}

\newcommand{\s}{{s$^{-1}$}}
\newcommand{\kms}{km\,s$^{-1}$}

\newcommand{\ciii}{\ion{C}{3}]}

\newcommand{\civ}{\ion{C}{4}}
\newcommand{\lya}{Ly$\alpha$}
\newcommand{\heii}{\ion{He}{2}}

\newcommand{\oiii}{\ion{O}{3}]}
\newcommand{\foiii}{[\ion{O}{3}]}

\newcommand{\ha}{H$\alpha$}
\newcommand{\hb}{H$\beta$}

\newcommand{\tigm}{$T_\mathrm{IGM}$}
\newcommand{\bubble}{GS-z6IS}

\newcommand{\editone}[1]{\textbf{#1}}
\renewcommand{\editone}[1]{{#1}} 
\graphicspath{{figure/}}

\begin{document}

\title{Using \lya\ Transmitted Spectrum to Probe IGM Transmission and Identify Ionized Structures in Cosmic Reionization}

\correspondingauthor{Weida Hu}
\email{weidahu@tamu.edu}

\author[0000-0003-3424-3230]{Weida Hu}
\email{weidahu@tamu.edu}
\affiliation{Department of Physics and Astronomy, Texas A\&M University, College Station, TX 77843-4242, USA}
\affiliation{George P. and Cynthia Woods Mitchell Institute for Fundamental Physics and Astronomy, Texas A\&M University, College Station, TX 77843-4242, USA}

\author[0000-0001-7503-8482]{Casey Papovich}
\email{papovich@tamu.edu}
\affiliation{Department of Physics and Astronomy, Texas A\&M University, College Station, TX 77843-4242, USA}
\affiliation{George P. and Cynthia Woods Mitchell Institute for Fundamental Physics and Astronomy, Texas A\&M University, College Station, TX 77843-4242, USA}

\author[0000-0001-9495-7759]{Lu Shen}
\email{lushen@tamu.edu}
\affiliation{Department of Physics and Astronomy, Texas A\&M University, College Station, TX 77843-4242, USA}
\affiliation{George P. and Cynthia Woods Mitchell Institute for Fundamental Physics and Astronomy, Texas A\&M University, College Station, TX 77843-4242, USA}

\author[0000-0002-7959-8783]{Pablo Arrabal Haro}
\email{parrabalh@gmail.com}
\affiliation{Center for Space Sciences and Technology, UMBC, 5523 Research Park Dr, Baltimore, MD 21228 USA }
\affiliation{Astrophysics Science Division, NASA Goddard Space Flight Center, 8800 Greenbelt Rd, Greenbelt, MD 20771, USA}

\author[0000-0001-8534-7502]{Bren E. Backhaus}
\email{bren.backhaus@ku.edu}
\affil{Department of Physics and Astronomy, University of Kansas, Lawrence, KS 66045, USA}

\author[0000-0001-7151-009X]{Nikko J. Cleri}
\email{cleri@psu.edu}
\affiliation{Department of Astronomy and Astrophysics, The Pennsylvania State University, University Park, PA 16802, USA}
\affiliation{Institute for Computational and Data Sciences, The Pennsylvania State University, University Park, PA 16802, USA}
\affiliation{Institute for Gravitation and the Cosmos, The Pennsylvania State University, University Park, PA 16802, USA}

\author[0000-0001-5414-5131]{Mark Dickinson}
\affiliation{NSF’s National Optical-Infrared Astronomy Research Laboratory, 950 N. Cherry Ave., Tucson, AZ 85719, USA}
\email{}

\author[0000-0002-1404-5950]{James S. Dunlop}
\email{james.dunlop@ed.ac.uk}
\affiliation{Institute for Astronomy, University of Edinburgh, Royal Observatory, Edinburgh EH9 3HJ, UK}

\author[0000-0001-8519-1130]{Steven L. Finkelstein}
\affiliation{Department of Astronomy, The University of Texas at Austin, Austin, TX, USA}
\email{stevenf@astro.as.utexas.edu}

\author[0000-0002-7831-8751]{Mauro Giavalisco}
\affiliation{University of Massachusetts Amherst, 710 North Pleasant Street, Amherst, MA 01003-9305, USA}
\email{}

\author[0000-0002-6610-2048]{Anton M. Koekemoer}
\email{koekemoer@stsci.edu}
\affiliation{Space Telescope Science Institute, 3700 San Martin Drive, Baltimore, MD 21218, USA}

\author[0000-0001-8519-1130]{Vasily Kokorev}
\affiliation{Department of Astronomy, The University of Texas at Austin, Austin, TX, USA}
\email{}

\author[0000-0002-9572-7813]{Sara Mascia}
\email{}
\affiliation{INAF - Osservatorio Astronomico di Roma, Via Frascati 33, 00078 Monte Porzio Catone, Rome, Italy}
\affiliation{Institute of Science and Technology Austria (ISTA), Am Campus 1, A-3400 Klosterneuburg, Austria }

\author[0000-0002-8951-4408]{Lorenzo Napolitano}
\email{lorenzo.napolitano@inaf.it}
\affiliation{INAF - Osservatorio Astronomico di Roma, Via Frascati 33, 00078 Monte Porzio Catone, Rome, Italy}
\affiliation{Dipartimento di Fisica, Università di Roma Sapienza, Città Universitaria di Roma - Sapienza, Piazzale Aldo Moro, 2, 00185, Roma, Italy}

\author[0000-0001-8940-6768]{Laura Pentericci}
\email{laura.pentericci@inaf.it}
\affiliation{INAF - Osservatorio Astronomico di Roma, Via Frascati 33, 00078 Monte Porzio Catone, Rome, Italy}

\author[0000-0002-2838-9033]{Aaron Smith}
\email{asmith@utdallas.edu}
\affiliation{Department of Physics, The University of Texas at Dallas, Richardson, Texas 75080, USA}

\author[0000-0003-1282-7454]{Anthony J. Taylor}
\email{anthony.taylor@austin.utexas.edu}
\affiliation{Department of Astronomy, The University of Texas at Austin, Austin, TX, USA}
\affiliation{Cosmic Frontier Center, The University of Texas at Austin, Austin, TX, USA}

\author[0000-0002-9373-3865]{Xin Wang}
\email{xwang@ucas.ac.cn}
\affiliation{School of Astronomy and Space Science, University of Chinese Academy of Sciences (UCAS), Beijing 100049, China}
\affiliation{National Astronomical Observatories, Chinese Academy of Sciences, Beijing 100101, China}
\affiliation{Institute for Frontiers in Astronomy and Astrophysics, Beijing Normal University, Beijing 102206, China}

\author[0000-0003-3466-035X]{{L. Y. Aaron} {Yung}}
\email{yung@stsci.edu}
\affiliation{Space Telescope Science Institute, 3700 San Martin Drive, Baltimore, MD 21218, USA}

\begin{abstract}
We present a study of intergalactic medium (IGM) transmission at $4.5 < z < 6.5$ using high-signal-to-noise JWST/NIRSpec prism spectroscopy of 143 galaxies at $5<z<7$ from the CAPERS and JADES surveys. 
By comparing the observed flux blueward of \lya\ emission line to the prediction of spectral energy distribution modeling, we directly measure the IGM transmission along the individual galaxy sightlines. The average transmission measured from these galaxy sightlines is consistent with previous measurements based on luminous quasars.
Current NIRSpec spectroscopy is sufficiently deep to probe IGM transmission on single sightlines.
We find evidence for a highly ionized structure, \bubble, at $z\sim 5.75-6$ in the GOODS-S field based on the analysis of a high-S/N spectrum of one galaxy, GS-18846, at $z=6.335$. 
The IGM transmission of \bubble\ is $0.17\pm0.02$, an order of magnitude higher than the average of previous measurements at this redshift.
This structure has a line-of-sight scale of $\sim110$ cMpc and spatially extends over at least $21\times17$ cMpc$^2$.
\bubble\ is associated with a known large-scale galaxy overdensity at the same redshift, whose member galaxies show enhanced \lya\ visibility and a broader \lya\ equivalent width distribution compared to field galaxies at similar redshift. 
\editone{This result supports the interpretation that \lya\ overdensity can trace bubbles of increased IGM transmission, although environmental effects on galaxy properties
may also contribute.
Our study demonstrates that high-S/N galaxy spectra offer a powerful new approach to tracing ionized structures during the epoch of reionization.}
\end{abstract}


\section{Introduction} 

The reionization of neutral hydrogen in the intergalactic medium (IGM) is a milestone in the history of the Universe. 
During this epoch, the ionizing photons generated from the first stars, galaxies, and active galactic nuclei (AGNs) start impacting the surrounding IGM, transitioning the IGM from a neutral to a mostly ionized state \citep[e.g.,][]{Barkana2001,Fan2006a}. 
There has been much progress in narrowing down the timeline of cosmic reionization: measurement of the Thompson optical depth by the polarization of cosmic microwave background (CMB) constrains that the cosmic reionization occurs before $z\sim7.7\pm0.7$ \citep{PlanckCollaboration2020}; observations of quasars show that the cosmic reionization ends at $z\sim5$ -- 6 \citep[e.g.,][]{Fan2006b,Becker2015,Eilers2018,Yang2020,Zhu2021}; statistic studies of \lya\ emitting galaxies (LAEs) and \lya\ break galaxies (LBGs) also show that the neutral hydrogen fraction in IGM decreases rapidly from $z>8$ to 5 \citep[e.g.,][]{Malhotra2004,Tilvi2014,Ouchi2010,Zheng2017,Mason2018,Hu2019,Morales2021,Wold2022,Bolan2022,Ning2022,Tang2024,Napolitano2024}.

One of the most direct and model-independent methods to constrain cosmic reionization is by measuring the IGM transmission from the transmitted flux in the region between rest-frame wavelength the \lya\ and Ly$\beta$ transitions (rest-frame $\lambda_0=1027$ -- 1216 \AA; we refer to this region as the ``\lya\ transmitted spectrum'' hereafter).
Neutral hydrogen in the IGM resonantly scatters these redshifted photons, strongly suppressing the transmitted intensity and even resulting in complete absorption (i.e., the Gunn-Peterson trough; \citealp{Gunn1965}) when the neutral hydrogen fraction is $<0.01\%$.
Thanks to the extreme luminosities of high-redshift quasars, this method has been effectively applied to the quasar spectra over the last three decades, providing precise measurements of IGM transmission at $5\lesssim z\lesssim 6$ \citep{Fan2006a,Becker2015,Eilers2018,Yang2020,Bosman2022}.
The differences in IGM transmission between the quasar sightlines at a fixed redshift also reveal that reionization does not proceed isotropically, but rather follows a patchy topology \citep{Becker2015}.
However, two problems occur when using quasars.  
\editone{First, the sparse distribution of quasars across the sky limits our ability to
sample the three-dimensional structure of the nearby IGM.}
Second, quasars are bright themselves, prone to ionizing large regions around them \citep[this is the well-known ``proximity effect'';][]{Faucher-Giguere2008}.  

Galaxies, though much fainter, are far more numerous and densely distributed, allowing us to map the local patchiness of the IGM and directly trace how reionization proceeds across different environments.
\citet{Thomas2017} conducted the first such test using galaxy spectra from the VLT/VIMOS Ultra Deep Survey \citep{LeFevre2015} and measured the IGM transmission up to $z\sim5$ \citep[see also][]{Thomas2020,Thomas2021}. 
More recently, \citet{Kakiichi2023} utilized ultra-deep narrowband imaging to probe \lya\ transmission at $z\sim4.9$ using background galaxies and cross-correlated it with the distribution of LAEs at the same redshift. 
However, the strong atmospheric OH emission in the near-infrared and the rapidly increasing IGM attenuation at $z>5$ make it extremely challenging to push IGM transmission studies to higher redshifts.
The advent of \textit{JWST}, with its high sensitivity and freedom from atmospheric emission, now enables IGM transmission studies to reach into the reionization era. 
\citet{Meyer2025} demonstrated this capability by combining NIRSpec/prism spectra of galaxies at $5<z<11$, finding IGM transmission measurements consistent with those derived from quasar sightlines.

In this work, we present a study of IGM transmission using \textit{JWST}/NIRSpec prism spectra of UV-bright galaxies from the CAPERS and JADES surveys and explore high signal-to-noise (S/N) galaxy spectra to probe ionized structures.
In Section \ref{sec:data}, we describe the data used in this work. 
Section \ref{sec:igm} outlines our method for measuring IGM transmission and presents the results.
We identify three high-transmission regions at $z\gtrsim5.3$ using the individual high S/N galaxy spectra, including a region at $z\sim5.9$ previously reported in \citet{Meyer2025}.
In Section \ref{sec:hightans}, we confirm this region to be a large ionized structure at $z\sim5.9$ in the GS field associated with a galaxy overdensity, and then explore \lya\ emission of galaxies within this structure.
Finally, Section \ref{sec:summary} summarizes our results.

Throughout this study, we adopt the Planck 2018 cosmological parameters \citep{PlanckCollaboration2020}: $\Omega_m=0.3111$, $\Omega_\Lambda=0.6889$ and $H_0=67.66$ \kms\ Mpc$^{-1}$, where $\Omega_m$ and $\Omega_\Lambda$ are the densities of total matter and dark energy and $H_0$ is the Hubble constant.

\section{Data and Sample} \label{sec:data}

The objects investigated in this work are drawn from the \textit{JWST} CANDELS-Area Prism Epoch of Reionization Survey (CAPERS) and \textit{JWST} Advanced Deep Extragalactic Survey (JADES). 
In Section \ref{sec:capers} and \ref{sec:jades}, we briefly introduce the observation and data reduction for CAPERS and JADES, respectively. 
In Section \ref{sec:sample}, we describe the selection of the galaxy sample.

\subsection{CAPERS} \label{sec:capers}

CAPERS is targeting seven NIRSpec pointings in each of the EGS, COSMOS, and UDS fields (21 fields in total) using the  PRISM/CLEAR spectroscopic element which provides low resolution, $R\sim 100$, wavelength coverage from 0.6 to 5.3 $\mu$m.
For each pointing, CAPERS used three NIRSpec MSA configurations with the lower-priority objects being replaced in each observation. 
Therefore, this setup enables high-priority objects to be observed multiple times to achieve a high S/N, while observing a large sample of lower-priority objects to maximize scientific yield.
Each MSA configuration is carried out using a three-shutter slitlet with nodding between the three shutters.
The effective exposure time of each MSA configuration is 5,690 s. Targets may be observed on one to three MSA configurations, depending on their priority, such that total effective exposure time per target ranges from 5,609 s to 17,079 s.
At the time of this study, CAPERS had completed the observations in the EGS and COSMOS fields, as well as 38\% of the planned observations for the UDS.
Therefore, this work focuses on CAPERS observations from the available data in these fields. 

The NIRSpec data were reduced using the \textit{JWST} Calibration Pipeline v1.17.1 \citep[][]{bushouse2025} with several customized steps similar to \citet{ArrabalHaro2023}.
A detailed description of data reduction will be presented in a dedicated paper.
The 1D spectra are then extracted using an optimal extraction \citep{horne1986}.
The galaxies' redshifts are measured using a variety of software, including a modified version of Marz, a modified version of MSAEXP, Bagpipes, Cigale, etc. \citep{Hinton2016Marz,Brammer2023,Carnall2018,Boquien2019}. 
The results are then combined and vetted by the CAPERS team.

The photometric catalogs in the three fields are obtained from the UNICORN project (S. Finkelstein et al., in prep), which provides robust photometry measurements (including accurate colors, total fluxes, and noise estimates) in a similar method as used in \citet{Finkelstein2024}.  
The IDs of CAPERS galaxies are from the UNICORN catalog.

\subsection{JADES} \label{sec:jades}

JADES \citep{Eisenstein2023} is a joint program between the \textit{JWST} NIRCam and NIRSpec GTO teams, targeting the Great Observatories Origins Deep Survey \citep[GOODS;][]{Giavalisco2004} South and North fields (GS and GN, respectively).
The program includes multi-wavelength imaging with NIRCam and MIRI, along with low- and medium-resolution spectroscopy with NIRSpec.
The MSA configurations are also designed to be a three-shutter slitlet with nodding between the three shutters.
In this work, we use the low-resolution spectroscopy obtained with the NIRSpec disperser/filter configuration of PRISM/CLEAR (R $\sim100$) from the JADES DR3.
The exposure times of JADES spectra span a wide range, from $\sim0.9$ hrs to as much as $\sim62$ hrs.
The JADES NIRSpec data were retrieved from the MAST HLSP archive\footnote{\url{https://archive.stsci.edu/hlsp/jades}}. 
We refer the readers to \citet{Bunker2024} for details.

We utilize the published \textit{JWST} images from the JADES DR3 \citep{Eisenstein2023,Eisenstein2023b,D'Eugenio2024}.
JADES NIRCam observation provides 14 broad- or medium-bands (broad- and medium-band imaging from F090W, F115W, F150W, F182M, etc. filters from JADES, \citealp{Hainline2024}, combined with medium-band imaging in F430M, F460M, and F480M from JEMS, \citealp{Williams2023}) covering wavelengths from 0.8 to 5.0 $\mu$m.
The JADES photometric catalog provides matched-aperture flux densities for the galaxies, and photometric redshifts measured from the SEDs \citep{Rieke2023,Eisenstein2023b,Hainline2024}.
In addition to the \textit{JWST} images, we also utilize the archival Hubble Space Telescope (HST) images to complement the \textit{JWST} images in the UV-to-optical wavelength range.
The archival HST images were obtained from several HST observation programs, including CANDELS, HUDF, UVCANDELS, etc \citep{Grogin2011,Koekemoer2011,Wang2025,Bouwens2011}.
The IDs of JADES galaxies are from the JADES NIRSpec IDs.

\subsection{Sample Selection} \label{sec:sample}

Our goal is to study the \lya\ transmission of IGM using the spectra of individual galaxies at redshifts near the reionization.
To this end, we select the galaxies using the following criteria:
(1) The wavelength range between \lya\ (rest-frame 1215.7 \AA) and Ly$\beta$ (rest-frame 1025.7 \AA) of the galaxies is covered by the NIRSpec PRISM spectra (0.6 -- 5.3 $\mu$m; corresponding to $z>5$);
(2) The galaxies are subject to relatively small IGM attenuation. We adopt $z<7$, where IGM transmission is $\sim0.0001$ predicted by the theoretical model \citep{Inoue2014};
(3) The rest-UV continua at 1400 -- 1600 \AA\ of galaxies are detected with a median S/N $>5$ per wavelength pixel;
(4) The galaxies have imaging observations with NIRCam F090W, F115W, F150W, F200W, F277W, F356W, and F444W broadbands to improve the SED fitting.

\begin{figure}
    \centering
    \includegraphics[width=1\linewidth]{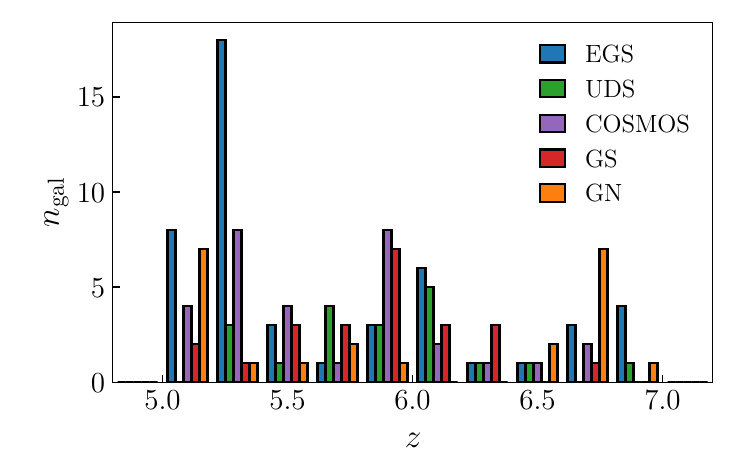}
    \caption{Histogram of redshifts of the galaxy sample used for the IGM transmission measurements.
    The blue, green, purple, red, and orange correspond to galaxies from the EGS, EDS, GS, and GN fields.}
    \label{fig:zdist}
\end{figure}

As the background of the CAPERS and JADES data is subtracted using the nodding pattern, any neighboring objects covered by the MSA shutters can result in oversubtraction of the background.
Therefore, we visually inspect the individual spectra of galaxies and their broadband images to remove those contaminated by the nearby objects.
We also remove the little red dots based on the morphologies and broad Balmer emission lines, as it is hard to model their spectra with the standard SED templates.
However, we note that an unknown fraction of AGNs may still be present in our sample.
In total, we select 143 galaxies, where 48, 19, 31, 23, and 22 galaxies are in EGS, UDS, COSMOS, GS, and GN fields, respectively. 
In Figure \ref{fig:zdist}, we present the redshift distribution of the galaxies.

\section{IGM Transmission} \label{sec:igm}

To measure the IGM transmission, we need to normalize the observed spectrum by an intrinsic, IGM-unattenuated spectrum appropriate for each galaxy. 
To achieve this, for each galaxy we perform a joint fitting to the galaxy's spectral energy distribution (SED) using all available photometry and the NIRSpec/PRISM spectrum to obtain a model  spectrum. We describe our method in Section \ref{sec:sed} and then describe the IGM transmission measurements in Sections \ref{sec:igmmeas} and \ref{sec:igmgalaxy}.

\subsection{SED fitting} \label{sec:sed}

We use the Python package \texttt{BAGPIPES} \citep{Carnall2018} to jointly fit the photometry and spectra of the objects.
We employ Binary Population and Spectral Synthesis v2.2.1 \citep[BPASS;][]{Eldridge2017,Stanway2018} with a broken power-law initial mass function with slopes of $\alpha_1 = -1.3$ for stars with 0.1 -- 0.5 $M_\odot$ and $\alpha_2 = -2.35$ for 0.5 -- 100 $M_\odot$ (model `135\_100').
For the star formation history parameterization, we use the Gaussian Process model from \texttt{dense\_basis} \citep{Iyer2019}, where the star formation history is split into 4 dynamically adjusted time bins, and during each time bin, $25\%$ of the total stellar mass is formed.
The stellar metallicity is allowed to vary between 0.001 and 1~$Z_\odot$.  
We include nebular emission with the metallicity of the gas equal to that of the stellar populations, and an ionization parameter,  $\log U$, in the range $-4$ to $-1$. 

The slope of the UV continuum and its extrapolation into the wavelength range blueward of \lya\ are sensitive to the choice of dust attenuation law.
The Calzetti dust law, derived from local star-forming galaxies \citep{Calzetti2000}, has been frequently applied in high-redshift studies. However, recent analysis \citep[e.g.,][]{Markov2025} have revealed that the dust law of high-redshift galaxies may deviate from this empirical dust law, potentially introducing systematic biases in the estimation of IGM transmission.
To account for this uncertainty, we adopt a flexible modification of the Calzetti dust law, as formulated by \citet{Salim2018}.
This parametrization has two free parameters: $\delta$, quantifying the deviation from the slope of the Calzetti law, and $B$, representing the strength of the 2175 \AA\ absorption feature.
When $\delta=0$, the Salim dust law is equivalent to the Calzetti dust law.
In this study, we fix $B=0$, as the wavelength range of interest is minimally affected by this feature.
We allow $A_V$ and $\delta$ to vary in the range of 0 -- 8 and -1.6 -- 0.4, respectively.

To account for the slitlosses and calibration uncertainties of NIRSpec spectroscopy, we adopt a multiplicative scaling factor (assumed to be a second-order Chebyshev polynomial) in \texttt{BAGPIPES} to estimate the wavelength-dependent flux calibration to the broadband photometry.
We allow the zeroth order to vary from 0.1 to 10, while the first and second orders are allowed to vary from $-0.5$ to 0.5. 
To account for the variable resolution of the NIRSpec spectrum, we utilize the resolution model for the prism spectra from the \textit{JWST} User Documentation\footnote{\url{https://jwst-docs.stsci.edu/jwst-near-infrared-spectrograph/nirspec-instrumentation/nirspec-dispersers-and-filters}}.
We also allow the emission lines to be broadened by an intrinsic velocity dispersion ranging from 0 to 500 \kms, though this is always much smaller than the instrument dispersion.
\editone{We note that a velocity dispersion of 0 \kms\ is physically implausible; this serves only as a mathematical boundary for the parameter space. In practice, the posteriors of the velocity dispersion typically range from 50 to 200 km/s.}
For redshifts, we use the internal redshift catalog of the CAPERS and the publicly released redshift catalog from JADES, allowing a small varying range of $\Delta z = \pm$0.05 to account for uncertainties.
For other parameters, we use the uniform priors across the allowed range.

\texttt{BAGPIPES} adopts the IGM attenuation model of \citet{Inoue2014} as default to account for the IGM attenuation. 
\citet{Inoue2014} derived the distribution functions of intergalactic absorbers from the observational statistics of Lyman Limit systems, \lya\ Forest systems, and Damped \lya\ systems, and then integrated the distribution function to derive the mean transmission function of IGM. 
However, as shown in the quasar studies, the IGM transmission contains a large scatter, and the scatter increases with increasing redshift (for example, the effect optical depth ranges from $\sim3$ to 6 at $5.7<z<5.9$ in \citealp{Yang2020}; see also \citealp{Becker2015,Eilers2018}).
Including the IGM attenuation model in the SED fitting could bias our estimation of IGM transmission. 
To avoid this, in the SED fitting, we mask the wavelength range blueward of the \lya\ break.
Additionally, since the \lya\ line cannot be reliably modeled with the SED-fitting codes and the \lya\ break could be affected by strong Damped \lya\ absorption \citep{Hu2023,Heintz2024,Heintz2025}, we mask both the \lya\ emission line and the \lya\ break region (rest-frame $<1270$ \AA).

\begin{figure}
    \centering
    \includegraphics[width=\linewidth]{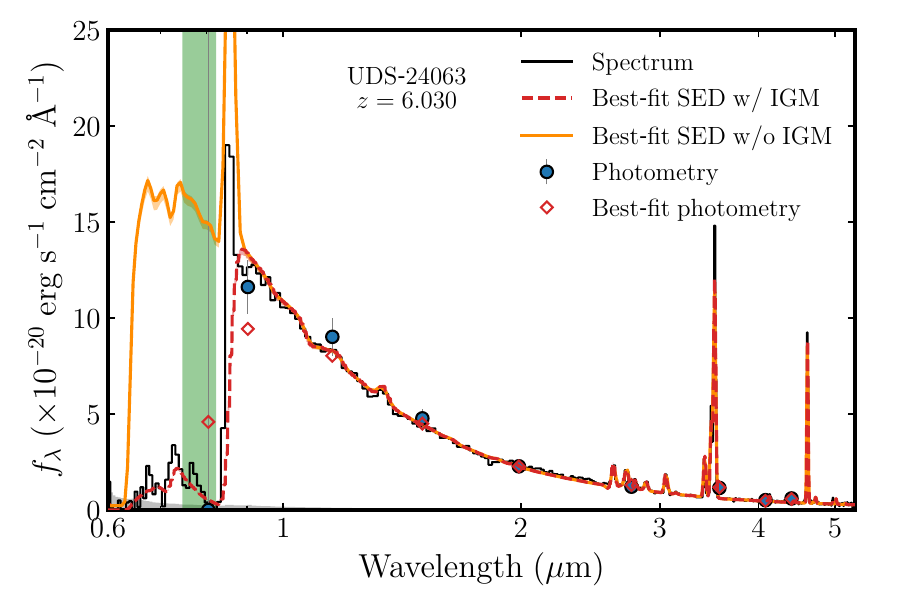}
    \caption{An example of SED fitting and IGM transmission measurements using the spectrum of CAPERS-UDS-24063. 
    The black line represents the observed spectrum, the gray shaded region represents the error spectrum, and the red dots represent the observed photometry. We present the IGM-unattenuated spectrum as the orange solid line and the IGM-attenuated spectrum (based on \citealp{Inoue2014} model) as the red dashed line. 
    The uncertainties of the IGM-unattenuated model and the IGM-attenuated model are shown as the orange- and red-shaded regions.
    The orange diamonds show the model flux densities in the imaging bands. We use a green shaded region to highlight the wavelength range used to calculate the IGM transmission.}
    \label{fig:uds24063}
\end{figure}

In Figure \ref{fig:uds24063}, we present the best-fit SED model of galaxy CAPERS-UDS-24063 as an example. Both the modeled spectra with and without the IGM attenuation are shown for comparison.
\editone{Because we require the selected galaxies to have UV continua detected at $>5\sigma$ per wavelength pixel and model their SEDs by jointly fitting spectra and photometry, the uncertainty in the predicted galaxy continua is very small with $\sigma(f_\mathrm{pred})/f_\mathrm{pred}<0.15$ in the wavelength range of \lya\ transmission.}

\subsection{IGM Transmission Measurements} \label{sec:igmmeas}

\begin{figure}
    \centering
    \includegraphics[width=1\linewidth]{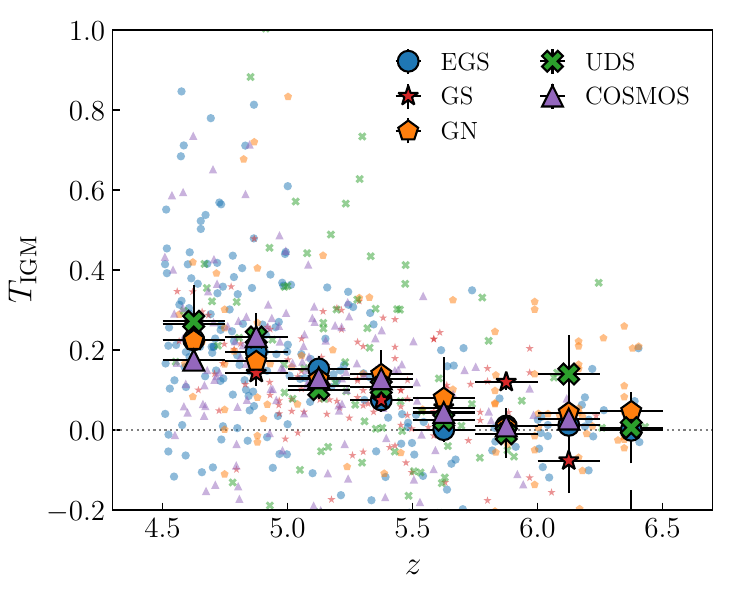}
    \caption{IGM transmission in EGS (blue circle), UDS (green cross), COSMOS (purple triangle), GS (red star), and GN (orange pentagon) fields. We present the measurements along individual galaxy sightlines as small, semi-transparent symbols.
    Large solid symbols represent the average transmissions in each field. 
    The horizontal error bars indicate the redshift bin widths ($\Delta z=0.25$) used to calculate the average transmission, and the vertical error bars indicate the measurement errors on the weighted average transmission.
    The error bars for individual measurements are excluded for clarity.
    The gray dotted line indicates $T_\mathrm{IGM}=0$, which is the lower limit of the IGM transmission.}
    \label{fig:igmtrans_field}
\end{figure}

\begin{figure*}[htbp]
    \centering
    \includegraphics[width=\linewidth]{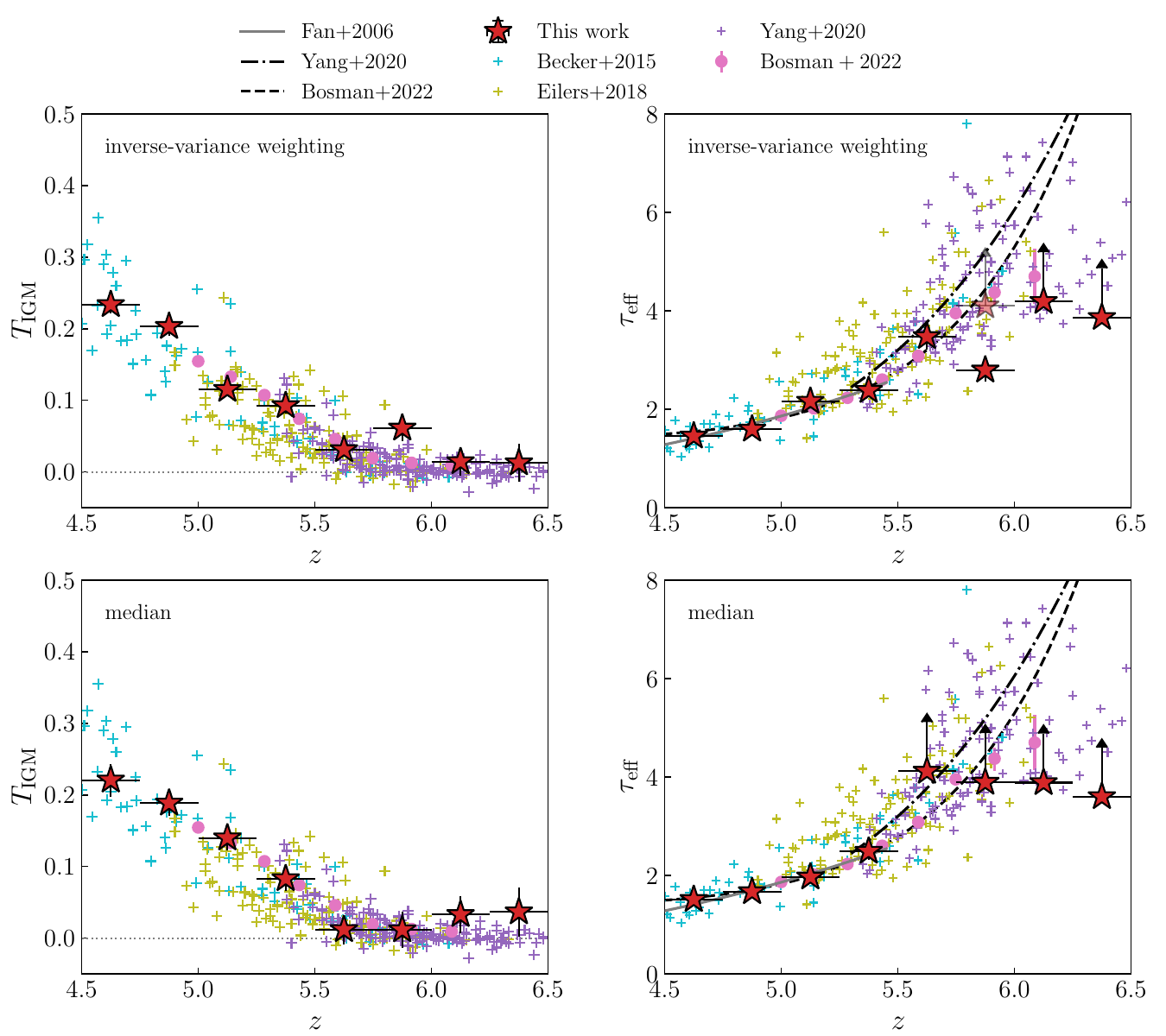}
    \caption{Top left: Comparison of the inverse-variance weighted IGM \lya\ transmission measured from the galaxies in the five fields to those from the literature. \editone{We plot the literature measurements of individual quasar sightlines as smaller crosses \citep{Becker2015,Eilers2018,Yang2020} and the average measurements as the pink dots \citep{Bosman2022}.}
    The gray dotted line indicates $T_\mathrm{IGM}=0$, which is the lower limit of the IGM transmission.
    The horizontal error bars indicate the redshift bin widths ($\Delta z=0.25$) used to calculate the average transmission, and the vertical error bars indicate the measurement uncertainties.
    The dispersion of the measurements is not included in the error bars.
    Top right: The \lya\ optical depth of IGM measured from the galaxies in the five fields.
    The red semi-transparent star presents the measurement of \lya\ optical depth in the other four fields after excluding GS.
    We plot the measurements from the literature \citep{Becker2015, Eilers2018,Yang2020,Bosman2022}. The gray solid line represents the best fit of optical depth evolution $\tau \propto (1+z)^{4.3}$ at $z<5.5$ \citep{Fan2006a}, the black dash-dotted line represents the best fit of optical depth evolution $\tau \propto (1+z)^{8.6}$ at $z>5.3$ \citep{Yang2020}, \editone{and the black dashed line represents the best fit of optical depth evolution $\tau= 0.30((1+z)/5.8)^{13.7}+1.35$ at $4.8<z<5.9$ \citep{Bosman2022}. Bottom left and right: Same as the top panels, but for IGM transmission and optical depth derived by the median-stacking method.}}
    \label{fig:igmtrans}
\end{figure*}

We derive the \lya\ transmission by normalizing the observed \lya\ transmitted spectrum to the predicted IGM-unattenuated spectra, $T = f_\mathrm{obs}/f_\mathrm{pred}$. 
\editone{The resolution element of the prism spectra corresponds to a physical scale of 44 -- 56 cMpc $h^{-1}$ for \lya\ transmission at $z\sim$ 4.5 -- 6.
To match this resolution, we integrate the \lya\ transmission within the fixed bin size of 50 cMpc $h^{-1}$ to derive an effective \lya\ transmission. 
This bin size is also consistent with those adopted in previous studies \citep{Becker2015,Eilers2018,Yang2020,Bosman2022}. 
Therefore, the effective \lya\ transmission, \tigm, along the line of sight of each galaxy is determined by:}
\begin{equation} \label{eq:1}
    T_\mathrm{IGM} = \int_{50\ \mathrm{cMpc}\ h^{-1}}  T(D_c) d D_c/(50\ \mathrm{cMpc}\ h^{-1}), 
\end{equation}
where $D_c$ is the comoving distance along the line of sight.
Due to the low spectral resolution of the prism spectra ($R\sim$ 35 -- 50 at 6000 -- 9000 \AA), we restrict the calculation of \tigm\ to the restframe wavelength range of 1060 -- 1170 \AA\ to avoid contamination from the \lya\ and Ly$\beta$ breaks.
\editone{The uncertainty in the IGM transmission is calculated by propagating the uncertainties in the observed and modeled spectra, although the contribution from the modeled spectrum is negligible compared to the observational uncertainties.}

To improve the S/N, we combine the \lya\ transmission from all sightlines. 
We compute the average transmission over the redshift range 4.5 -- 6.5 using a bin size of $\Delta z = 0.25$. 
Within each redshift bin, we adopt inverse-variance weighting to average the \lya\ transmission and the associated uncertainties.
In Figure \ref{fig:igmtrans_field}, we present the \lya\ transmission measured from individual galaxy sightlines and the average \lya\ transmissions of the five fields.
The average \lya\ transmissions of the five fields agree within $1\sigma$ uncertainties.
\editone{We note that the inverse-variance weighting method could be biased towards high S/N sightlines, such as CAPERS-UDS-24063. We therefore also measure IGM transmission using a median-stacking method.
The median-stacking method yields larger uncertainties than the inverse-variance weighting, and at $z>5.5$ it provides only $1\sigma$ upper limits.
In the top-left and bottom-left panels of Figure \ref{fig:igmtrans}, we present the combined \lya\ transmission derived using the inverse-variance weighting and median stacking, respectively.}
We present our measurements in Table \ref{tab:igm}.

\editone{Because we are probing very weak signals that are typically undetectable in the spectrum of an individual galaxy, even small systematic effects, such as background residuals, could bias our measurements. 
It is therefore important to verify that our uncertainties are not affected by systematic offsets. To test this, we examine the distribution of noise-normalized fluxes in spectral regions where no intrinsic galaxy flux is expected.
Specifically, we select galaxies at $z>7.5$ and extract spectral pixels in the wavelength range 7000 -- 7500 \AA, corresponding to rest-frame wavelengths beyond the Lyman limit (rest-frame 912 \AA). 
Because the CAPERS and JADES data are reduced using different pipelines, we analyze the two datasets separately, yielding 46 galaxies in the CAPERS sample and 24 galaxies in the JADES sample.
For each sample, the distribution of noise-normalized fluxes (i.e., S/N per spectral pixel) is well described by a Gaussian and is consistent with a standard normal distribution (mean $=0$, $\sigma=1$). 
Therefore, the spectral regions with low flux level are dominated by random noise rather than systematic biases.}

We calculate the effective optical depth, $\tau_\mathrm{eff}$, by:
\begin{equation} \label{eq:2}
    \tau_\mathrm{eff} = -\ln\ T_\mathrm{IGM}.
\end{equation}
The IGM transmission at $z>6$ is detected only at $\sim 1\sigma$ significance; we adopt the $1\sigma$ uncertainties of \tigm\ to derive conservative lower limits on $\tau_\mathrm{eff}$.
The results are shown in the top-right and bottom-right panels of Figure \ref{fig:igmtrans}.
The IGM transmissions (optical depths) derived in this study decrease (increase) with increasing redshift, consistent with our expectation that the neutral hydrogen fraction increases with increasing redshift.

In Figure \ref{fig:igmtrans}, we compare our measurements with those from the previous quasar studies \citep{Becker2015,Eilers2018,Yang2020,Bosman2022}, along with the best-fit relations between average IGM optical depth and redshift reported in literature \citep{Fan2006b,Yang2020,Bosman2022}.
At $z<5.75$, our measurements of IGM transmission from the two methods are consistent with quasar-based studies within the $1\sigma$ scatter.
\editone{However, at $5.75<z<6.00$, our optical depth derived by the inverse-variance weighting is significantly lower than the best-fit relations derived in both \citet{Yang2020} and \citet{Bosman2022}, though it is not seen in the median-stacking results.}
This deviation is predominantly driven by the high $T_\mathrm{IGM}$ observed with high S/N in the GS field, as shown in Figure \ref{fig:igmtrans_field}.
If we exclude the GS field, we derive a lower limit on the effective optical depth $\tau_\mathrm{eff}>4.1$ of the other 4 fields, agreeing with the previous studies and the average trend.
We interpret this as physical field-to-field variance in the IGM transmission.  We discuss the excess in the GS field further below in Section \ref{sec:hightans}.

\subsection{High Transmission Regions along Individual Sightlines} \label{sec:igmgalaxy}

\begin{figure}
    \centering
    \includegraphics[width=\linewidth]{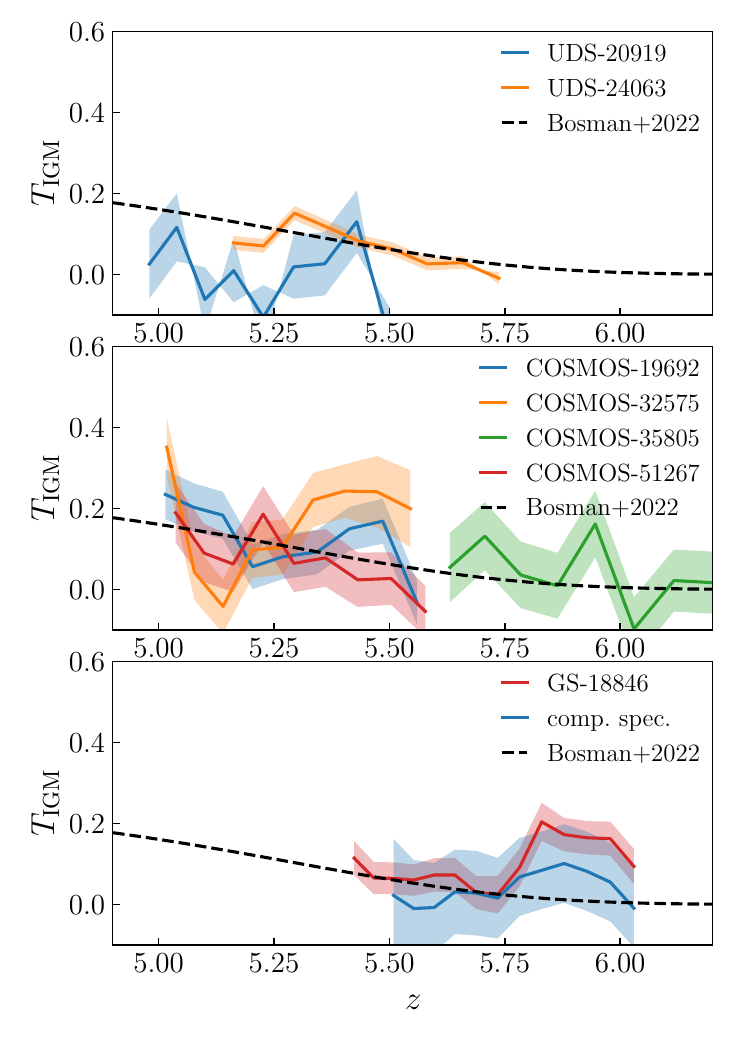}
    \caption{IGM transmission along individual sightlines of high S/N galaxies. From top to bottom panels, we present the measurements in the UDS, COSMOS, and GS fields. The shaded regions represent the $1\sigma$ error.
    In each panel, we present the prediction of the empirical relation in \citet{Bosman2022} for comparison. 
    In the bottom panel, we also present the average IGM transmission measured from the galaxies at $6.3<z<7$ in GS, excluding GS-18846.}
    \label{fig:hightrans}
\end{figure}

Several spectra in our sample have sufficient S/N to allow accurate direct measurements of IGM transmission along individual galaxy sightlines.
As our primary goal is to probe IGM in the epoch of reionization ($z>5.5$), we focus on galaxies at $z>5.8$.
To ensure reliable measurements, we require a median continuum S/N $>15$ per wavelength interval in the rest-frame 1400 -- 1600 \AA\ continuum, yielding 7 galaxies from our sample.
In Figure \ref{fig:hightrans}, we present their IGM transmission spectra, grouped by field.
While most of these individual measurements are consistent with both previous studies and \editone{the empirical relation of \citet{Bosman2022}}, three galaxies show evidence of higher transmission compared to the prediction of the empirical relation over an extended redshift range: UDS-24063, COSMOS-32575, and GS-18846. We discuss each of these below.

\vspace{1em}
\textit{UDS-24063:} UDS-24063 at $z=6.030$ provides the highest-S/N UV continuum among our sample (shown in Figure \ref{fig:uds24063}), with a median of S/N $\sim64$ per wavelength pixel in the rest-frame wavelength range of 1400 -- 1600 \AA. 
The top panel of Figure \ref{fig:hightrans} shows that its IGM transmission agrees well with the predictions of the empirical relation at $z>5.5$ and $z<5.25$ within $1\sigma$, but at $5.25<z<5.5$ the IGM transmission is higher.
Following Equation~(\ref{eq:1}), we measure the average IGM transmission $T_\mathrm{IGM}=0.11\pm0.01$ in this redshift range, compared to the prediction of the empirical relation, 0.086, a significance of $2.4\sigma$. This excess falls at the upper end of measurements from quasar sightlines, which range from 0.00 to 0.14 at $5.25<z<5.5$ \citep{Eilers2018}. 
We note that this high transmission is not shown in the sightline of another galaxy, UDS-20919, at a projected distance of 130 arcsec, though its IGM transmission measurement shows a large uncertainty.

\vspace{1em}
\textit{COSMOS-32575:} COSMOS-32575 at $z=5.841$ shows an excess transmission with an average $T_\mathrm{IGM}=0.23\pm0.04$ at $5.3<z<5.55$, as presented in the middle panel of Figure \ref{fig:hightrans}. This is larger than the prediction (0.076) of the empirical relation at $\sim4\sigma$ significance.
However, two additional galaxies in the COSMOS field do not show comparable transmission excess at the same redshift:
COSMOS-19692 shows a modest excess at a narrower redshift range 5.35 -- 5.5, with an average $T_\mathrm{IGM}=0.16\pm0.04$, and COSMOS-51267 is fully consistent with the empirical relation.
We note that the projected separations of COSMOS-19692 and COSMOS-51267 from COSMOS-32575 are 251 arcsec and 375 arcsec, corresponding to 9.9 cMpc and 14.8 cMpc. 
This variation of transmission across nearby sightlines could reflect the geometry of the high transmission region. 
Future higher spectral resolution observations will be essential to reconstruct its detailed geometry.

\vspace{1em}
\textit{GS-18846}: GS-18846 at $z=6.335$ shows elevated IGM transmission  at $5.75<z<6$ with an average $T_\mathrm{IGM}=0.17\pm0.02$.  This is an order of magnitude higher than the prediction of the empirical relation or $T_\mathrm{IGM}=0.012$. This redshift range corresponds to a path length of 73 cMpc h$^{-1}$ = 109 cMpc (for the adopted cosmology).

GS-18846 is the only galaxy in the GS field with sufficient S/N to probe its individual sightline, unlike the situation in the COSMOS field, where several such galaxies exist. 
To investigate the IGM transmission of nearby sightlines, we average the spectra of several galaxies at redshifts $6.3 < z < 7$ that have detections of their rest-frame UV continua at 1400--1600~\AA\ at $>1\sigma$ significance (excluding GS-18846). 
We choose this redshift range to avoid contamination from the \lya\ break and the Ly$\beta$ forest, and we also removed 1 galaxy that is likely contaminated by a nearby object, yielding a sample of 22 galaxies.
We apply the same method presented in Sections \ref{sec:sed} and \ref{sec:igmmeas} to measure the IGM transmission for these individual sightlines, and we then average the results weighting by the inverse variance. 
The bottom panel of Figure \ref{fig:hightrans} shows this average transmission, which has tentative evidence for an excess IGM transmission with an integrated $T_\mathrm{IGM}=0.073\pm0.039$ over $5.75 < z < 6$, consistent with the increased transmission seen in GS-18846. 
This reveals that the high-transmission region could extend across the entire GS spectroscopic footprint ($\sim$ 21 cMpc $\times$ 17 cMpc).

\begin{figure}[tbp]
    \centering
    \includegraphics[width=\linewidth]{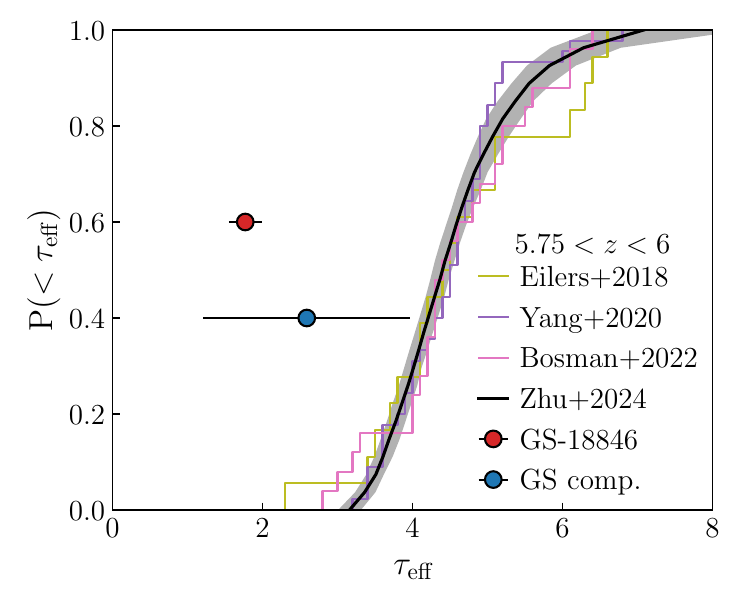}
    \caption{Comparison between the optical depths at $5.75<z<6$ probed by galaxies in GS and those probed by quasars. The red and blue dots represent our measurements from GS-18846 and the average transmission. The purple and yellow curves represent the CDFs obtained from the quasar spectra \citep{Eilers2018,Yang2020,Bosman2022}. The black curve is the CDF derived from the cosmological hydrodynamic simulation CROC \citep{Zhuh2024}, and the gray shaded region represents its scatter.  The IGM optical depths measured from GS-18846 and the average of other GS show lower optical depth than measured in all but one quasar sightlines (see text).}
    \label{fig:taudist}
\end{figure}

The persistently high transmission over such a long path length at $z\sim5.9$ is surprising and has not been previously observed except in quasar proximity zones.
Using Equation~(\ref{eq:2}), we derive the optical depth $\tau_\mathrm{eff}=1.77\pm0.12$ for GS-18846 and $\tau_\mathrm{eff}=2.59\pm0.53$ for the average transmission of other galaxies at $6.3<z<7$.
In Figure \ref{fig:taudist}, we compare the effective optical depths to the cumulative distribution of $\tau_\mathrm{eff}$ at $5.75<z<6$ derived from quasar spectra \citep{Eilers2018,Yang2020,Bosman2022}. 
Most quasar sightlines in this redshift range exhibit $\tau_\mathrm{eff}>3$ with a median of $\sim 4.5$, significantly larger than the optical depth of the sightline of GS-18846 and tentatively larger than the optical depth of average transmission.
Only one quasar sightline, J1306+0356, from \citet{Eilers2018} shows comparably low optical depth, with $\tau_\mathrm{eff} = 2.37 \pm 0.01$. 
However, this enhanced transmission in the sightline of J1306+0356 is confined to a region within $\sim$50 cMpc $h^{-1}$ of the quasar and is likely associated with the proximity zone of the quasar.
We note that the optical depths in \citet{Eilers2018}, \citet{Yang2020}, and \citet{Bosman2022} are measured over 50 cMpc $h^{-1}$ bins. 
Adopting a larger bin size, such as 73 cMpc $h^{-1}$ for GS-18846, would reduce the variance and increase the minimum optical depth in quasar studies.

The high IGM transmission suggests that the line of sight toward GS-18846 passes through a volume with a low neutral hydrogen fraction at $z\sim 5.9$, or alternatively a volume with a high ionized gas fraction. 
Interestingly, this corresponds to an overdensity of galaxies at this redshift.  We explore the properties of the galaxies in this high transmission structure in GS in Section \ref{sec:hightans}.

\section{A High Transmission Region at $5.75<\lowercase{z}<6$ in GS} \label{sec:hightans}

A high transmission region with a size of tens of cMpc can enable \lya\ photons being shifted out of resonance before they encounter neutral IGM, and therefore, can substantially reduce the IGM attenuation and enhance the \lya\ visibility of galaxies within the ionized region \citep{Dijkstra2014,Jung2022,Larson2022,Yue2025}.
This has led to the use of overdensity of \lya-emitting galaxies as tracers of ionized bubbles and reionization \citep{Tilvi2020,Hu2021,Endsley2021,Jung2022,Whitler2025,Chen2025}.  Here, we show that the high IGM transmission at $5.75 < z < 6$ in the sightline of GS-18446 and the average of other GS galaxies corresponds to a galaxy overdensity at this redshift.  
In Section \ref{sec:gs18446}, we first explore the origin of the high-transmission region in the sightline of GS-18846 and its connection with the underlying galaxies.
In Sections \ref{sec:gs9422} and \ref{sec:lyaew}, we examine the \lya\ properties of the galaxies in the high transmission region.   
Our conclusion is that this overdensity of \lya-emitters in the GS field traces an ionized bubble as indicated by the higher IGM transmission in this field.

\subsection{An ionized bubble \bubble\ associated with an overdensity} \label{sec:gs18446}

\begin{figure*}
    \centering
    \includegraphics[width=0.68\linewidth]{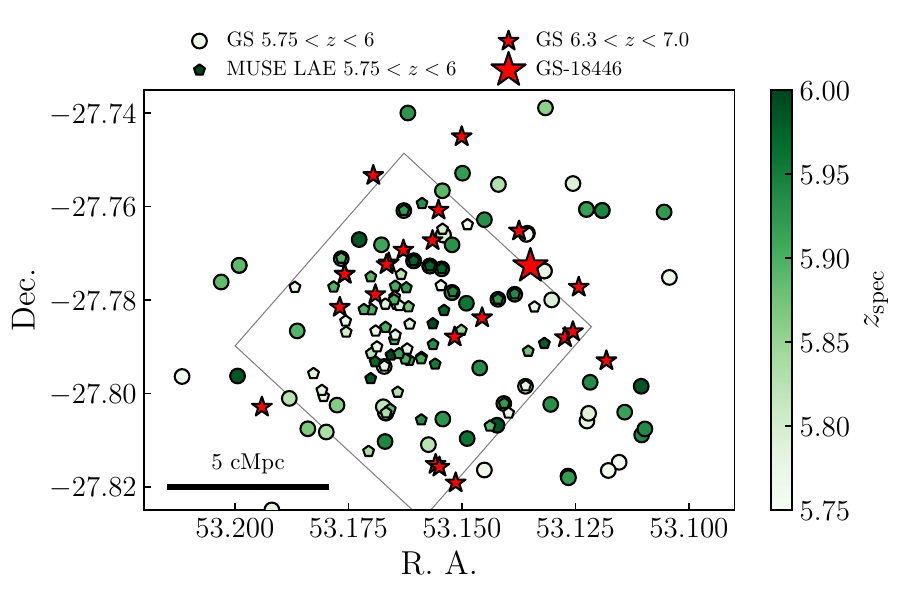}
    \includegraphics[width=0.28\linewidth]{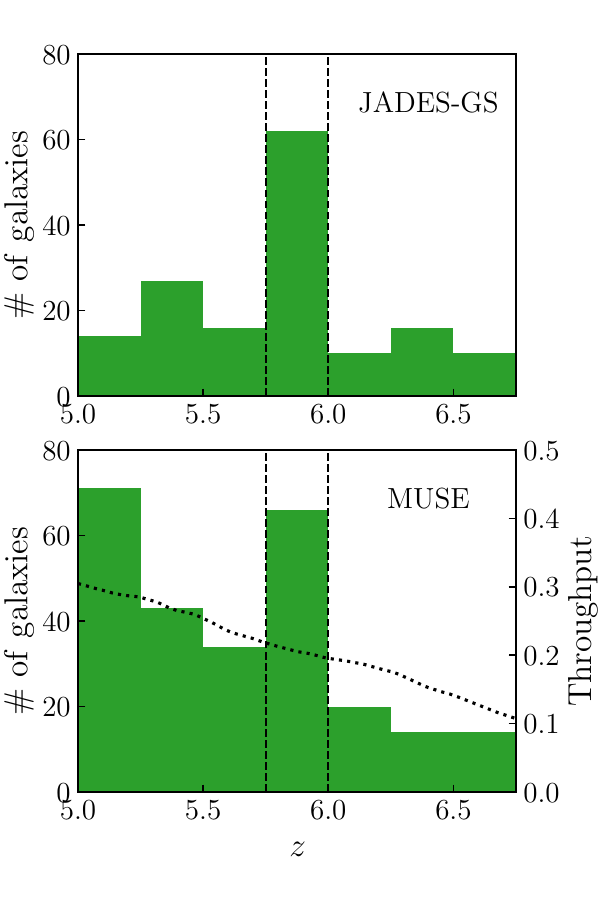}
    \caption{\textbf{Left:} Spatial distribution of \bubble\ member galaxies. The circles and pentagons represent the spectroscopically confirmed galaxies at $5.75<z<6$ from the JADES-GS data and the LAEs at $5.75<z<6$ from the MUSE HUDF catalog, respectively. We color-code the symbols by the redshifts of galaxies. 
    The gray rectangle illustrates the field of view of the MUSE HUDF survey.
    We also plot the background galaxies that are used to probe the IGM in \bubble. The large red star represents GS-18846, and the small red stars represent the galaxies at $6.3<z<7$ whose spectra are used to obtain average IGM transmission. 
    In Figure \ref{fig:sky3d}, we present a 3D illustration.
    \textbf{Right:} Redshift distributions of spectroscopically confirmed galaxies at $5<z<6.75$ from the JADES-GS catalog and LAEs at $5<z<6.75$ from the MUSE HUDF catalog. The dashed lines highlight the redshift range of \bubble. The dotted curve in the bottom panel shows the total throughput of MUSE.}
    \label{fig:z6bubble}
\end{figure*}

\begin{figure*}
    \includegraphics[width=\linewidth]{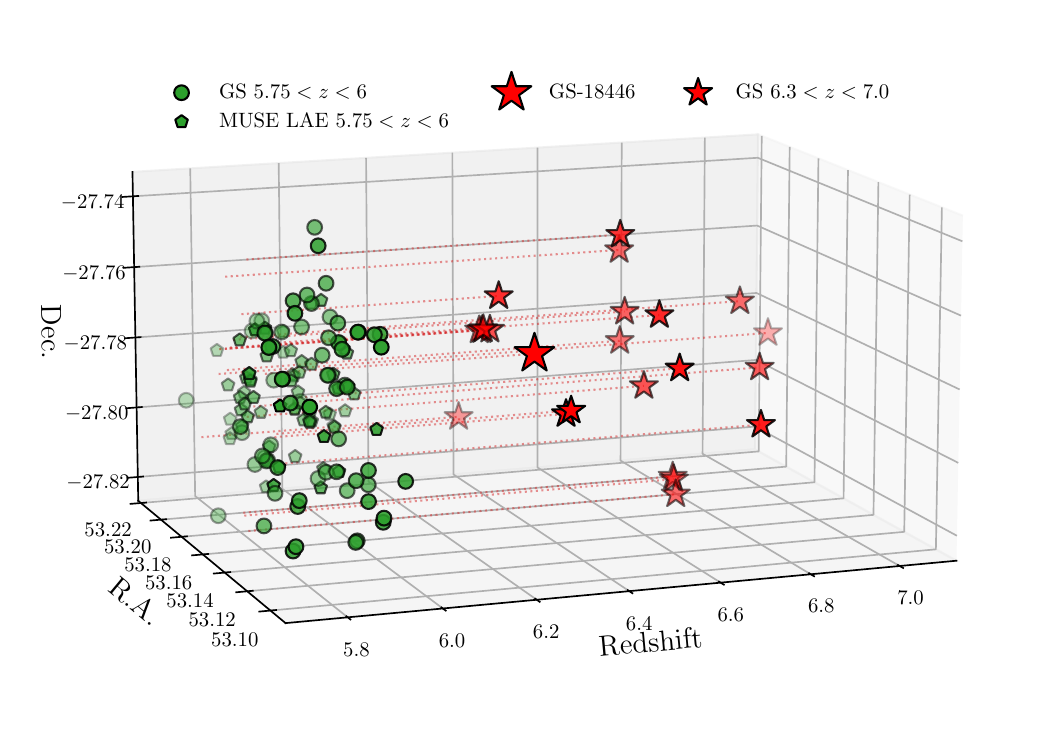}
    \caption{3D spatial distribution of \bubble\ galaxies (green) and background galaxies (red). 
    The green circles and pentagons represent the spectroscopically confirmed galaxies at $5.75<z<6$ from the JADES-GS data and the LAEs at $5.75<z<6$ from the MUSE HUDF catalog, respectively.
    The large red star represents GS-18846, and the small red stars represent the galaxies at $6.3<z<7$ whose spectra are used to obtain average IGM transmission. The red dotted lines illustrate how the sightlines of individual galaxies pass through \bubble. An interactive version of this figure will be available in the online journal.}
    \label{fig:sky3d}
\end{figure*}

The low neutral hydrogen density like that observed in GS-18846 at $5.75 < z < 6$ could be driven by an intrinsically underdense region--a large void of neutral IGM.  However, this case does not seem sufficient to produce the low optical depth observed for this galaxy.
Using the CROC cosmological hydrodynamic simulations, \citet{Zhuh2024} investigated the origin of \lya\ transmission spikes in quasar spectra and found that such spikes can be explained by the underdense regions.
To explore this possibility, we compare the optical depth along the sightline of GS-18846 with the CDF of optical depths at $5.75 < z < 6$ from the CROC simulation in Figure \ref{fig:taudist}. 
The CROC CDF is calculated based on integrating the \lya\ transmission along 45 simulated sightlines over this redshift range.
The CROC CDF agrees well with quasar-based measurements, but yields optical depths that are much larger than the optical depth measured in the spectrum of GS-18846.
This discrepancy suggests that a low IGM density alone cannot explain the unusually high transmission along this sightline.

Rather, the high \lya\ transmission is likely due to an increase in the number of ionizing photons impacting the IGM at $5.75 < z < 6$ in the GS field, leading to a higher ionized hydrogen fraction. This suggests the presence of an overdensity of ionizing sources at $z \sim 5.9$ in this field, producing a ``bubble''. Hereafter, we name this ionized structure at $z\sim 6$ in the GS field: \bubble.

Producing such a large ionized structure when the IGM was still highly neutral requires a large-scale overdensity of star-forming galaxies and/or AGNs to generate substantial ionizing photons, such as expected from galaxy overdensities \citep[e.g.,][]{Witstok2024,Martin2025}.
To identify such a structure, we first use the JADES-GS spectroscopic catalog.
The transverse scale of the NIRSpec field of JADES in GS is $\sim 21$ cMpc $\times$ 17 cMpc, much smaller than the line-of-sight extent of the apparent ionized structure (109 cMpc).  Given the uncertain location of the center of the galaxy overdensity, we select all galaxies with secure redshifts (flag $<$ D) in this field with no restriction on sky location.
The top right panel of Figure \ref{fig:z6bubble} shows the redshift distribution of the galaxies between $z=5$ and 6.75 in the JADES-GS field, binned in $\Delta z=0.25$. 
A clear spike can be seen at the redshift bin 5.75 -- 6 with 64 galaxies, exceeding the other redshift bins with 10 -- 27 galaxies.
This implies a galaxy overdensity at 5.75 -- 6 with an overdensity factor of $\delta\sim4$, after correcting for volume differences.
This suggests that \bubble\ is associated with an overdensity at $z\sim5.9$.

We note that JADES spectroscopy observation was optimized to observe the most distant galaxies \citep{Bunker2024,D'Eugenio2024}. 
The complex selection functions could bias the redshift distribution of targets.
Particularly, the galaxies at $5.7<z<8$ were assigned slightly higher priority than the galaxies $4<z<5.7$. 
However, we do not expect the observed spike arising from these selection effects, given the high significance of the excess compared to that from the higher and lower redshift galaxies. 
Furthermore, the JADES-GN spectroscopy observation was designed similarly, and it shows no such comparable redshift overdensity at $5.75<z<6$.

We also explore for any evidence of a galaxy overdensity at this redshift in the GS field using archival data from the VLT/MUSE Hubble Ultra Deep field (HUDF) survey \citep{Bacon2017}.
The MUSE HUDF survey is an integral field spectroscopy survey of the HUDF field, covering the wavelength range from 4650 -- 9300 \AA.
The high-redshift galaxies in the MUSE HUDF survey are primarily identified through their \lya\ emission lines, i.e., LAEs. 
In the bottom right panel of Figure \ref{fig:z6bubble}, we present the redshift distribution of MUSE HUDF LAEs.
The number of LAEs gradually decreases with the redshift due to the decreasing instrument sensitivity for longer wavelengths and increasing IGM attenuation for higher redshifts.
Despite this, a peak remains at $5.75<z<6$, with twice the number of LAEs compared to the declining trend, supporting the presence of an overdensity.

Indeed, several independent studies over the past two decades have consistently reported evidence for an overdensity at $z\sim5.7$ -- 6.0 in the GS field.
\citet{Stanway2004} confirmed 3 LAEs with Gemini/GMOS-South at $z=5.79$, 5.83, and 5.94, and first suggested a possible overdensity at $z\sim5.8$ in the GS field.
Based on the ground-based, wide-field ($31' \times 33'$) narrowband imaging in the GS field, \citet{Wang2005} also identified an overdensity at $z\sim5.7$ -- 5.8 with 17 LAEs.
In addition, \citet{Malhotra2005} analyzed the HST/ACS grism spectroscopic observation of the GS field and identified an overdensity at $z=5.9\pm0.2$ consisting of 15 Lyman Break galaxies, with an overdensity factor of 4. 
More recently, \citet{Helton2024} reported an overdensity at $z\sim5.93$ with $\delta=4.923^{+0.583}_{-1.038}$ in the GS field based on the NIRCam WFSS observation from the FRESCO survey \citep{Oesch2023}. \citet{D'Eugenio2025} reported an overdensity at $z\sim6$ with a $3\sigma$ significance based on the NIRCam photometric catalog.    

Together, these observations strongly support our interpretation that the ionized structure we identify at $z\sim5.9$ is associated with a large-scale galaxy overdensity in the GS field.
We note that this ionized structure was also recently independently reported by \citet{Meyer2025}, who demonstrated that it can be explained by ionizing photons produced by galaxies down to M$_\mathrm{UV}<-10$ in the overdensity, assuming an ionizing photon escape fraction of $40\%$ and an ionizing photon production efficiency of $\log_{10}\ \xi_\mathrm{ion}=25.4$.

\begin{figure}
    \centering
    \includegraphics[width=1\linewidth]{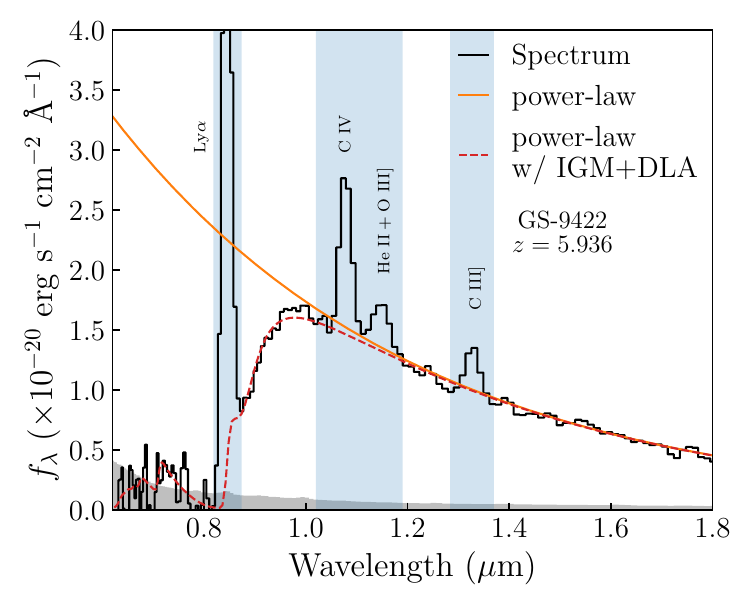}
    \caption{NIRSpec Prism spectrum of GS-9422 and the best-fit model using a power-law function accounting for a \textit{partial-covering} DLA and the IGM attenuation. The black line and the gray-shaded region represent the observed spectrum and the error spectrum. The blue shaded regions mask the \lya, \civ, \heii+\oiii, and \ciii\ emission lines that are excluded in the fitting.
    We present the best-fit power-law function to GS-9422 as the orange solid line and the best-fit model accounting for the DLA and IGM attenuation as the red dashed line.}
    \label{fig:gs9422}
\end{figure}

\subsection{GS-9422: A Galaxy with Most of Its Escaped \lya\ Photons Observed?} \label{sec:gs9422}

Among the member galaxies of \bubble, we first examine an extreme \lya\ emitting galaxy at $z=5.936$, GS-9422, which shows strong UV and optical high-ionization lines and a remarkable rest-frame UV continuum turnover at 1400 \AA.
The 1400 \AA\ turnover could be explained as caused by the partial-covering damped \lya\ absorber system (DLA) or the two-photon emission from transition 2s $\rightarrow$ 1s in hydrogen \citep{Cameron2024,Tacchella2025}.
In this section, we argue that the 1400 \AA\ turnover is caused by a \textit{partial-covering} damped \lya\ absorber based on its \lya\ transmitted spectrum, and we argue that the escaped \lya\ photons of GS-9422 experience very little IGM attenuation.  

The high-ionization UV lines of GS-9422, such as \civ, \heii+\oiii, and \ciii, suggest a highly ionized ISM, which is not readily modeled by the standard SED fitting packages. 
However, the extra-galactic objects, such as AGNs or star-forming galaxies, usually exhibit a power-law UV continuum \citep[e.g.,][]{Meurer1999,Dunlop2013,Cai2023,Tacchella2025}. 
Therefore, we alternatively adopt a power law ($f_\lambda\propto \lambda^\beta$) to model the UV continuum.
We mask the emission lines, including \lya, \civ, \heii+\oiii, and \ciii, to avoid their contribution to the continuum fitting.
We model the 1400 \AA\ turnover using a modified Voigt profile, which is defined by a Doppler parameter, a neutral hydrogen column density $N_\mathrm{HI}$, a velocity offset of the DLA system, and a covering factor $f_c$, similar to \citet{Hu2023}.
Because the DLA absorption of GS-9422 is very wide and the spectral resolution is considerably low ($R\sim40$), we fix the velocity offset to be 0 and adopt a typical Doppler parameter 30 \kms\ \citep{Prochaska2015}. 
We also consider the IGM attenuation based on the \citet{Inoue2014} model.
Furthermore, we utilize the Python package \texttt{SpectRes} \citep{Carnall2017} to account for the resolution model for the prism spectra. 
We use a Markov Chain Monte Carlo sampler \texttt{emcee} \citep{Foreman-Mackey2013} to compute the posterior of the parameters.
We obtain a column density of $10^{23.2\pm0.1}$ cm$^{-2}$ and a covering fraction of $0.65\pm0.06$ for the DLA. 
The best-fit model is presented in Figure \ref{fig:gs9422}.

Our model successfully reproduces the 1400 \AA\ turnover, aside from a minor residual excess at 1400 \AA.
We note that this excess could be contributed by the Si \textsc{iv} 1400 \AA\ emission line that is frequently shown in the quasar spectra. 
More importantly, our model also reproduces the transmitted spectrum between 7000 and 7700 \AA, which cannot be explained by the two-photon emission model in \citet{Cameron2024}.
Because in the two-photon emission model, the continuum blueward of \lya\ would be lower than the extrapolation of UV continuum in 1400 -- 1600 \AA\ by a factor of $>3$ (see their Figure 13), resulting in an underprediction of the transmitted spectrum by a factor of $>3$.
Therefore, the 1400 \AA\ turnover should be contributed by the combination of a power-law continuum and a \textit{partial-covering} DLA, and the AGN-dominated model proposed by \citet{Tacchella2025} provides a more physically plausible explanation for this origin.

The IGM transmission of \lya\ photons of a galaxy can be estimated by:
\begin{equation}
    T_\mathrm{IGM} = \frac{F^\mathrm{obs}_\mathrm{Ly\alpha}}{F^\mathrm{int}_\mathrm{Ly\alpha}\times f_\mathrm{esc}^\mathrm{Ly\alpha}},
\end{equation}
where $F^\mathrm{obs}_\mathrm{Ly\alpha}$ is the observed \lya\ flux, $f_\mathrm{esc}^\mathrm{Ly\alpha}$ is the \lya\ escape fraction ranging from 0 to 1, and $F^\mathrm{int}_\mathrm{Ly\alpha}$ is the intrinsic \lya\ flux which can be inferred from the dust-corrected \ha\ flux assuming a Case B recombination: $f^\mathrm{int}_\mathrm{Ly\alpha}=8.7\times F_\mathrm{H\alpha}$.
Because $f_\mathrm{esc}^\mathrm{Ly\alpha}$ is difficult to constrain, we instead calculate the total transmission:
\begin{equation} \label{eq:gs9422}
    T_\mathrm{total} = T_\mathrm{IGM}\times f_\mathrm{esc}^\mathrm{Ly\alpha} = \frac{F^\mathrm{obs}_\mathrm{Ly\alpha}}{8.7\times F_\mathrm{H\alpha}}.
\end{equation}
The total transmission $T_\mathrm{total}$ should range from 0 to 1, and a value of $T_\mathrm{total} \approx 1$ implies that both the ISM and IGM are largely transparent to \lya\ photons. 

\begin{deluxetable*}{l r}
    \tablecaption{Flux measurements of mentioned emission lines in this work for GS-9422 \label{tab:gs9422}}
    \tablehead{\colhead{Line} & \colhead{$f_\mathrm{line}$} \\ 
    \colhead{} & \colhead{$\times10^{-19}$ erg s$^{-1}$ cm$^{-2}$} }
    \startdata
    \lya & $127.5^{+2.5}_{-2.1}$ \\
    \civ & $39.8^{+1.8}_{-1.8}$ \\
    \heii+\oiii & $17.1^{+2.0}_{-1.9}$ \\
    \ciii & $11.4^{+1.3}_{-1.3}$ \\
    \hb & $18.68^{+0.26}_{-0.34}$ \\
    \foiii\ $\lambda4959$ & $34.00^{+0.36}_{-0.37}$ \\
    \foiii\ $\lambda5007$ & $104.29^{+0.57}_{-0.56}$ \\
    \ha & $49.90^{+0.36}_{-0.38}$ \\
    \enddata
\end{deluxetable*}

\editone{The extremely high column density of the partial-covering DLA around GS-9422 can produce saturated absorption, completely absorbing the photons at the \lya\ line center (rest-frame $\sim1215.67$ \AA).
Consequently, only \lya\ photons that escape through low-column density sightlines not covered by the DLA can traverse into the IGM and ultimately reach the observer \citep{Hu2023}. }
Therefore, we should consider a covering fraction correction $(1-f_c)$ in Equation~(\ref{eq:gs9422}):
\begin{equation} 
    T_\mathrm{total} = T_\mathrm{IGM}\times f_\mathrm{esc}^\mathrm{Ly\alpha} = \frac{F^\mathrm{obs}_\mathrm{Ly\alpha}}{8.7\times F_\mathrm{H\alpha}\times(1-f_c)}.
\end{equation}
We measure the \lya\ flux and Balmer line fluxes from the prism spectrum. 
To measure the \lya\ flux, we subtract the best-fit continuum model from the spectrum and then use a Gaussian profile to fit the emission line. 
The fluxes of Balmer lines are measured using a Gaussian model with a linear continuum model. The measurements are presented in Table \ref{tab:gs9422}, 
We also measure the \lya\ peak velocity offset relative to the spectroscopic redshift by fitting a Gaussian profile to the G140M grating spectrum, and obtain a $\Delta v=190\pm12$ \kms.
The \ha/\hb\ line flux ratio, $2.67^{+0.05}_{-0.05}$, is smaller than the theoretical prediction assuming the typical Case B recombination (2.86 when $T=10,000$ K and $n_e=100$ cm$^{-3}$; \citealp{Osterbrock2006}), suggesting no significant dust attenuation in GS-9422. Therefore, we do not consider the dust correction here. 
We then obtain a total transmission of $83\pm2\%$ from the \lya/\ha\ line ratio.
This indicates the IGM transmission $T_\mathrm{IGM}>83\pm2\%$ for a \lya\ line at 190 \kms.

Such a high IGM transmission revealed by the \lya\ emission line of GS-9422 also suggests the presence of a large ionized bubble. 
To estimate the bubble size, we adopt the prescription proposed in \citet{Mason2020}, which assumes that the galaxy resides in the center of an ionized bubble and the IGM is completely neutral outside the bubble. 
The neutral IGM outside the bubble and the residual neutral hydrogen within the bubble together contribute to the \lya\ attenuation.
Considering the recombination within the ionized bubble, the residual neutral hydrogen fraction is assumed to be $\chi_\mathrm{HI}\propto r^2$, where $\chi_\mathrm{HI}(0.1~\mathrm{pMpc})=10^{-8}$.
For the \lya\ line shape, we adopt a Delta function with a 190 \kms\ velocity offset.
For $T_\mathrm{IGM}>83\pm2\%$, we obtain a bubble size of $>34$ cMpc. 
This agrees with the length of the high transmission region observed in the spectrum of GS-18846, further supporting the presence of an ionized bubble at $5.75<z<6$.

\editone{We note that the unusually small \ha/\hb\ ratio in GS-9422 could indicate a deviation from the commonly assumed Case B recombination. \citet{Yanagisawa2024} proposed two scenarios to explain this deviation: (1) density-bounded nebulae with smaller optical depth than Case B recombination, and (2) ionization-bounded nebulae surrounded by optically thick excited H \textsc{i} clouds.
In the first scenario, high-order Lyman photons are converted into Balmer photons, resulting in a smaller \ha/\hb\ line ratio. In this case, the intrinsic \lya/\ha\ ratio should be smaller than the Caes B ratio of 8.7. Therefore, the \lya\ escape fraction would be larger, implying a larger bubble size.
In the second scenario, the \lya/\ha\ ratio should be larger than 8.7, because the optically thick excited H \textsc{i} clouds could attenuate \ha\ photons, meanwhile a large number of hydrogen atoms at the 2s state can enhance \lya\ emission. Therefore, the inferred bubble size would be smaller. 
However, as determining the physical origin of the Balmer decrement anomaly in GS-9422 is beyond the scope of this work, we adopt the Case B assumption to be consistent with the literature.
}

\subsection{\lya\ EWs of LAEs in \bubble} \label{sec:lyaew}

\begin{figure}[tbp]
    \centering
    \includegraphics[width=\linewidth]{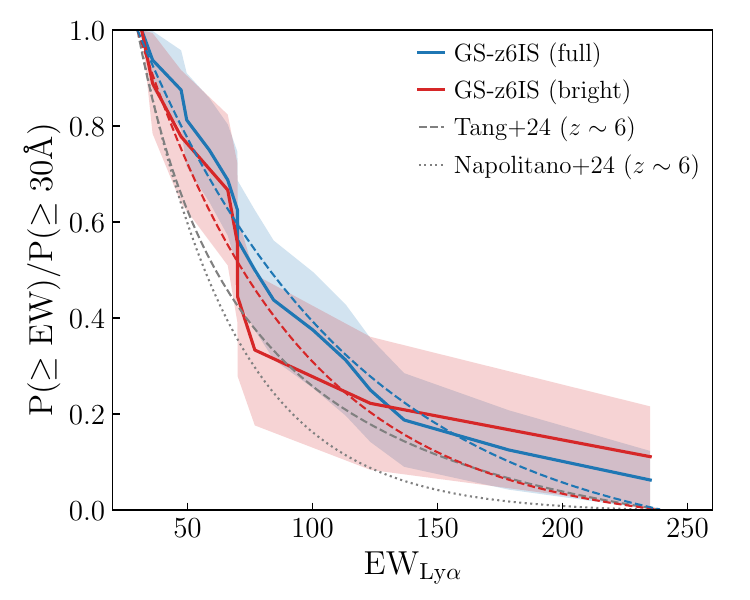}
    \caption{\lya\ EW CDF of LAEs in \bubble\ compared to field LAEs at $z\sim6$. We show the \lya\ EW CDF for LAEs in \bubble\ (red solid line), with the associated uncertainty as the red shaded region. 
    The best-fit exponential function to the \bubble\ CDF is overplotted as the purple dashed line.
    For comparison, we include \lya\ EW CDFs from previous studies.
    The gray dotted line shows the CDF of LAEs at $z<6$ in the \textit{JWST} CEERS and JADES surveys \citep{Napolitano2024}. The gray dashed line shows the CDF of LAEs at $z\sim6$ derived from a combination of \textit{JWST} and archival ground-based data \citep{Tang2024}.}
    \label{fig:lyaew}
\end{figure}

In this section, we explore the \lya\ EWs of spectroscopically confirmed galaxies in \bubble.
To robustly identify LAEs and characterize their \lya\ emission, we focus on the 46 galaxies with rest-frame UV continuum detected at $>1\sigma$ in the NIRSpec prism data.
We measure the \lya\ fluxes and EWs following the method applied to GS-9422 in Section \ref{sec:gs9422}.
While a partial-covering DLA absorption model was necessary for GS-9422, none of the remaining 45 galaxies show clear DLA features; therefore, we exclude this component from their spectral fits.
We then select LAEs as galaxies with \lya\ EW $>30$ \AA\ and S/N $>3$, consistent with the typical selection criteria in previous \textit{JWST} NIRSpec prism studies \citep[e.g.,][]{Jones2024b,Napolitano2024}.
In Table \ref{tab:1}, we present the galaxy properties and \lya\ measurements of the 46 galaxies. 
Based on this, we identify 16 LAEs among the 46 galaxies, corresponding to an LAE fraction of $X_\mathrm{Ly\alpha}\mathrm{(EW>30\AA)}=35^{+11}_{-9}\%$.
The uncertainty is derived from the small number statistic \citep{Gehrels1986}.

For a fair comparison with previous studies, which typically measure the LAE fraction from galaxies with $-20.25<\mathrm{M}_{UV}<-18.75$ for a high completeness, we focus on the 25 galaxies within this UV magnitude range and identify 9 LAEs.
This corresponds to an LAE fraction of $X_\mathrm{Ly\alpha}\mathrm{(EW>30\AA)}=36^{+17}_{-12}\%$.
This fraction is larger than the average LAE fractions of $X_\mathrm{Ly\alpha}\mathrm{(EW>25\AA)}\sim 15 - 25\%$ reported at $z\sim6$ in previous \textit{JWST} studies \citep{Jones2024b,Napolitano2024,Napolitano2025}, providing further evidence that the IGM attenuation in \bubble\ is lower compared to other field environments.

To further investigate the IGM attenuation, we study the \lya\ EW distribution of LAEs in \bubble, shown in Figure \ref{fig:lyaew}.
Uncertainties on the cumulative fraction are calculated from the variance of the binomial distribution, and we do not include individual EW measurement errors, as these are considerably smaller than the binomial variance. 
We fit the EW distribution for the LAEs in \bubble\ with an exponential function:
\begin{equation}
    \mathrm{P(EW)}\propto e^\mathrm{-EW/\omega_0},
\end{equation}
and derive a scale length of $\omega_0=87^{+14}_{-11}$ \AA.
Here, we apply a cut of $\mathrm{EW}=240$ \AA\ in the fitting, as this is the theoretical upper limit of \lya\ EW \citep{Charlot1993,Malhotra2002}.
We also fit the EW distribution of LAEs with $-20.25<\mathrm{M}_{UV}<-18.75$ to be consistent with previous studies, and derive a scale length of $64^{+17}_{-12}$ \AA. We present the best-fit exponential functions as the blue and red dashed lines in Figure \ref{fig:lyaew}.

We then compare the EW distribution of \bubble\ with previous works.
\citet{Napolitano2024} studied the LAEs with UV magnitude $-20.25<\mathrm{M_{UV}}<-18.75$ from the \textit{JWST} CEERS and JADES surveys.
They adopted an exponential function to fit the EW distribution of LAEs at $z<6$ and obtained a scale length of $\omega_0=39\pm2$.
\citet{Tang2024b} combined the \textit{JWST} observations with the archive ground-based VLT/MUSE and Keck/DEIMOS observations to study the \lya\ properties of galaxies with $-20.25<\mathrm{M_{UV}}<-18.75$ at redshift 5 -- 6.
They fitted the EW distribution with a lognormal distribution:
\begin{equation}
    \mathrm{P(EW)}\propto \frac{1}{\sqrt{2\pi\sigma} x} e^{-\frac{(\ln x-\mu)^2}{2\sigma^2}},
\end{equation}
where $\mu$ is the mean of the log-normal distribution and $\sigma$ is the standard deviation. 
For $z\sim6$ galaxies, they derived $e^\mu=8^{+4}_{-3}$ and $\sigma=1.85^{+0.42}_{-0.33}$.
In Figure \ref{fig:lyaew}, we present the best-fit functions from these two studies. 
The \lya\ EW distribution of \citet{Tang2024b} extends to higher EWs than \citet{Napolitano2024}. 
This is due to the fact that the \lya\ emission is more extended than the UV emission \citep{Steidel2011}, and could be missed by the much smaller apertures of \textit{JWST} NIRSpec observations \citep[see also][]{Napolitano2025}.
Compared with the results of \citet{Napolitano2024} and \citet{Tang2024b}, the \lya\ EW distribution of LAEs with $-20.25<\mathrm{M_{UV}}<-18.75$ in \bubble\ lies above their best-fit functions. 
The derived scale length $\omega_0$ ($64^{+17}_{-12}$ \AA) is also larger than that in \citet{Napolitano2024}, however, with a significance of only $2\sigma$ limited by the relatively small LAE sample size in \bubble.

We also compare the \lya\ EW distribution of \bubble\ with those pre-\textit{JWST} measurements.
These earlier studies reported a wide range of scale length $\omega_0$, caused by the differences in observational techniques and UV magnitude ranges of the samples.
For example, \citet{Hashimoto2017} used deep MUSE observation of HUDF field to study LAEs and derive scale lengths of $\omega_0=71\pm19$ and $\omega_0=81\pm36$ \AA\ for LAEs with $\mathrm{M_{UV}}<-18.5$ at $z\sim3.6$ and 4.9, respectively.
\citet{Zheng2014} measured $\omega_0=56$ \AA\ for the narrowband-selected LAEs at $z\sim4.5$.
\citet{Santos2020} used a set of narrow- and medium-band filters to select LAEs at $2<z<6$ and derived $\omega_0\sim 100$ -- 140 \AA\ for LAEs at $z\sim2.5$ -- 4.7 ($\omega_0\sim 80$ -- 110 \AA\ if applying an EW cut of 240 \AA).
We note again that these pre-\textit{JWST} studies typically suffer less from aperture loss than our \textit{JWST} NIRSpec measurements and adopt different UV magnitude ranges.
Despite these differences, the scale length of the \lya\ EW distribution in \bubble\ is broadly consistent with these pre-\textit{JWST} results. 

Taken together, our analysis shows that the LAEs in \bubble\ exhibit enhanced \lya\ EWs relative to the field galaxies at similar redshift, but are comparable to galaxies observed at later epochs when the IGM was fully ionized.
\editone{This provides additional evidence that galaxies in \bubble\ experience reduced IGM attenuation, consistent with the presence of a highly ionized bubble.
Nevertheless, as \bubble\ is associated with a galaxy overdensity and \lya\ escape is known to correlate with galaxy properties such as star formation rate and outflow kinematics \citep{Matthee2016,Yang2017}, environmental effects on galaxy properties may also contribute to the enhanced \lya\ EWs.}

\section{Summary} \label{sec:summary}

In this work, we present direct measurements of IGM transmission at $4.5 < z < 6.5$ using high-S/N JWST/NIRSpec prism spectroscopy of 143 
galaxies at $5<z<7$ from the CAPERS and JADES surveys, and compare our measurements with previous works.
We also explore the high transmission regions at $z\gtrsim 5.5$ revealed by the individual high-S/N spectra. 
The main results are summarized as follows:

\begin{itemize}
    \item By comparing the observed \lya\ transmitted spectrum to the prediction of SED modeling, we measure the IGM transmission and effective optical depth along individual galaxy sightlines. We use the inverse variance weighting method to average the transmission measurements and find that our measurements at $4.50<z<5.75$ are consistent with those reported in previous studies. We find that the effective optical depth at $5.75<z<6.00$ is significantly lower than previous studies, which is driven by an ionized structure at $z\sim5.75$ -- 6.00. At $z>6.00$, the current data is not deep enough to provide robust constraints.
    
    \item We identify an ionized structure \bubble\ at $z\sim 5.75$ -- 6.00 in the GS field, independently confirming the earlier findings in \citet{Meyer2025}. 
    We detect \bubble\ from the high-S/N spectrum of GS-18846 at $z=6.335$. This structure has a scale of 109 cMpc and an IGM transmission of $0.17\pm0.02$, an order of magnitude higher than the empirical relation at this redshift.
    The average IGM transmission of other galaxies at $6.3<z<7$ in GS (excluding GS-18846) shows a tentative transmission excess at $z\sim5.75$ -- 6, indicating \bubble\ spatially extends over at least $21\times17$ cMpc$^2$.
    \editone{Our direct measurement of the scale of GS-z6IS can serve as a useful comparison for reionization simulations, such as \citet{Lu2024,Neyer2025}.}

    \item \bubble\ is associated with a known overdensity of LAEs and LBGs discovered two decades ago \citep{Stanway2004,Wang2005,Malhotra2005}.
    This, for the first time, provides direct evidence that the overdensity of LAEs can trace the highly ionized structures in the epoch of reionization.

    \item We study the \lya\ transmitted spectrum and \lya\ emission line of an extreme galaxy GS-9422 within \bubble. We find that its \lya\ transmitted spectrum favors a \textit{partial covering} damped \lya\ absorption origin with $N_\mathrm{H}=10^{23.2\pm0.1}$ \cmsq\ and $f_c=0.65\pm0.06$ for its previously-reported 1400 \AA\ turnover, and it has a total \lya\ transmission $T_\mathrm{IGM}\times f^\mathrm{Ly\alpha}_\mathrm{esc}=83\pm2\%$, close to unity. Such a high total transmission suggests a large ionized bubble, consistent with the scale of \bubble\ inferred from the \lya\ transmitted spectrum of GS-18846.

    \item We further explore the \lya\ of member galaxies within the ionized structure \bubble. We find that these member galaxies show enhanced \lya\ visibility and a broader \lya\ equivalent width distribution than field galaxies at similar redshift, providing additional evidence that \bubble\ traces a highly ionized structure of the intergalactic medium.
\end{itemize}

Interestingly, the IGM transmission within \bubble, as probed by GS-18846, is still below unity. 
This suggests that residual neutral hydrogen remains within this ionized structure.
One possible explanation is that the sightline of GS-18846 passes through dense, self-shielded neutral hydrogen clouds, such as filaments in the cosmic web \citep{Chardin2018,Garaldi2022}. 
Alternatively, the structure may consist of multiple ionized bubbles that have not yet completely merged, leaving large neutral islands in between. 
Owing to the limited spectral resolution of the NIRSpec prism data, we are unable to distinguish between these two scenarios. 
Future medium- and high-resolution NIRSpec spectroscopy will be essential for resolving the small-scale kinematics and transmitted features necessary to discriminate between these possibilities.

\begin{acknowledgments}
We thank Hanjue Zhu for kindly sharing the IGM transmission measurements from the CROC simulation.
This work is based on observations made with the NASA/ESA/CSA \textit{James Webb Space Telescope}, obtained at the Space Telescope Science Institute, which is operated by the Association of Universities for Research in Astronomy, Incorporated, under NASA contract NAS5-03127. Support for program number GO-6368 was provided through a grant from the STScI under NASA contract NAS5-03127. The data were obtained from the Mikulski Archive for Space Telescopes (MAST) at the Space Telescope Science Institute. The CAPERS observations are associated with program 6368 and can be accessed via DOI: 10.17909/0q3p-sp24
The JADES observations are associated with programs 1180, 1181, 1210, 1286, 1895, 1963, and 3215, and can be accessed via DOI: 10.17909/8tdj-8n28.
We thank Eduardo Ba\~{n}ados for providing access to the NIRCam F090W data for the CEERS EGS field (GO-2234).
\end{acknowledgments}


\facilities{\textit{JWST}}

\software{Astropy \citep{astropy:2013, astropy:2018, astropy:2022}, BAGPIPES \citep{Carnall2018}}

\begin{appendix}

\section{Best-fit SED Model of GS-18846}

\begin{figure}
    \centering
    \includegraphics[width=0.5\linewidth]{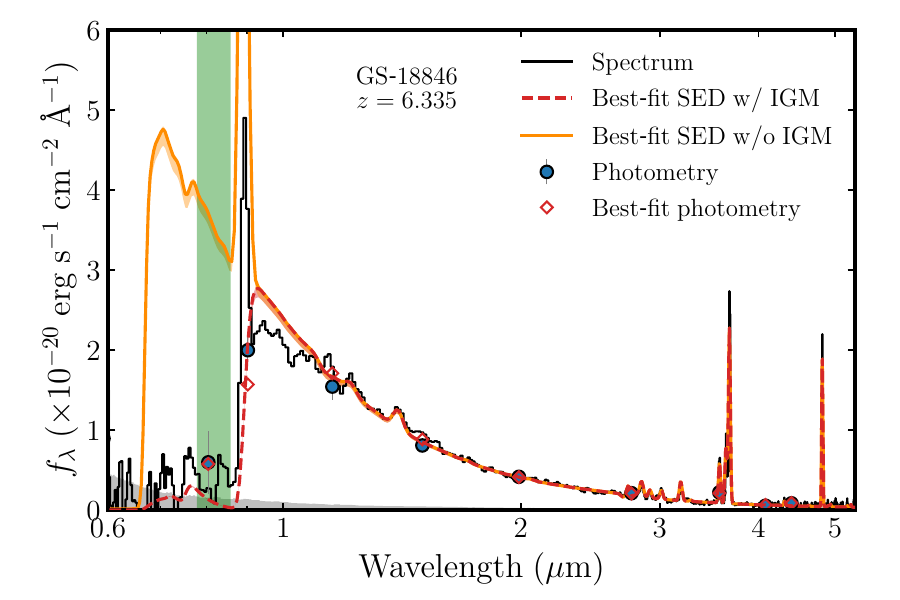}
    \caption{Same as Figure \ref{fig:uds24063}, but for GS-18846.}
    \label{fig:gs18846}
\end{figure}

\editone{
In Figure \ref{fig:gs18846}, we present the best-fit SED model of GS-18846 to show the excess of \lya\ transmission at $5.75<z<6$.}

\startlongtable
\begin{deluxetable}{c c c c}
    \tablecaption{IGM Transmission \label{tab:igm}}
    \tablehead{
    \colhead{Field} & \colhead{Redshift} & \colhead{$T_\mathrm{IGM}$} & \colhead{$\sigma(T_\mathrm{IGM})$}
    }
    \startdata
    \hline
    EGS & 4.50 -- 4.75 & 0.225 & 0.014 \\
     & 4.75 -- 5.00 & 0.196 & 0.014 \\
     & 5.00 -- 5.25 & 0.153 & 0.027 \\
     & 5.25 -- 5.50 & 0.075 & 0.024 \\
     & 5.50 -- 5.75 & 0.001 & 0.025 \\
     & 5.75 -- 6.00 & 0.010 & 0.030 \\
     & 6.00 -- 6.25 & 0.012 & 0.014 \\
     & 6.25 -- 6.50 & -0.001 & 0.021 \\
    \hline
    UDS & 4.50 -- 4.75 & 0.272 & 0.072 \\
     & 4.75 -- 5.00 & 0.232 & 0.049 \\
     & 5.00 -- 5.25 & 0.101 & 0.009 \\
     & 5.25 -- 5.50 & 0.107 & 0.011 \\
     & 5.50 -- 5.75 & 0.024 & 0.008 \\
     & 5.75 -- 6.00 & -0.010 & 0.048 \\
     & 6.00 -- 6.25 & 0.140 & 0.077 \\
     & 6.25 -- 6.50 & 0.006 & 0.071 \\
    \hline
    COSMOS & 4.50 -- 4.75 & 0.174 & 0.020 \\
     & 4.75 -- 5.00 & 0.234 & 0.012 \\
     & 5.00 -- 5.25 & 0.130 & 0.011 \\
     & 5.25 -- 5.50 & 0.128 & 0.032 \\
     & 5.50 -- 5.75 & 0.044 & 0.022 \\
     & 5.75 -- 6.00 & 0.010 & 0.035 \\
     & 6.00 -- 6.25 & 0.027 & 0.050 \\
     & 6.25 -- 6.50 & -0.295 & 0.116 \\
    \hline
    GS & 4.50 -- 4.75 & 0.265 & 0.013 \\
     & 4.75 -- 5.00 & 0.142 & 0.026 \\
     & 5.00 -- 5.25 & 0.111 & 0.015 \\
     & 5.25 -- 5.50 & 0.075 & 0.012 \\
     & 5.50 -- 5.75 & 0.056 & 0.016 \\
     & 5.75 -- 6.00 & 0.120 & 0.016 \\
     & 6.00 -- 6.25 & -0.077 & 0.065 \\
     & 6.25 -- 6.50 & ... & ... \\
    \hline
    GN & 4.50 -- 4.75 & 0.224 & 0.031 \\
     & 4.75 -- 5.00 & 0.172 & 0.032 \\
     & 5.00 -- 5.25 & 0.127 & 0.039 \\
     & 5.25 -- 5.50 & 0.139 & 0.049 \\
     & 5.50 -- 5.75 & 0.080 & 0.083 \\
     & 5.75 -- 6.00 & 0.009 & 0.031 \\
     & 6.00 -- 6.25 & 0.044 & 0.044 \\
     & 6.25 -- 6.50 & 0.047 & 0.037 \\
    \hline
    Total & 4.50 -- 4.75 & 0.234 & 0.010 \\
    (inverse-variance & 4.75 -- 5.00 & 0.203 & 0.010 \\
    weighting) & 5.00 -- 5.25 & 0.115 & 0.008 \\
     & 5.25 -- 5.50 & 0.093 & 0.009 \\
     & 5.50 -- 5.75 & 0.031 & 0.008 \\
     & 5.75 -- 6.00 & 0.061 & 0.014 \\
     & 6.00 -- 6.25 & 0.014 & 0.015 \\
     & 6.25 -- 6.50 & 0.015 & 0.022 \\
    \hline
    Total & 4.50 -- 4.75 & 0.226 & 0.018 \\
    (median) & 4.75 -- 5.00 & 0.199 & 0.015 \\
     & 5.00 -- 5.25 & 0.140 & 0.014 \\
     & 5.25 -- 5.50 & 0.086 & 0.014 \\
     & 5.50 -- 5.75 & 0.012 & 0.016 \\
     & 5.75 -- 6.00 & 0.011 & 0.020 \\
     & 6.00 -- 6.25 & 0.029 & 0.020 \\
     & 6.25 -- 6.50 & 0.037 & 0.027 \\
    \hline
    \enddata
\end{deluxetable}

\startlongtable
\begin{deluxetable*}{c c c c c c c}
    \tablecaption{Galaxy Properties in \bubble \label{tab:1}}
    \tablehead{
    \colhead{ID} & \colhead{RA} & \colhead{DEC} & \colhead{redshift} & \colhead{M$_\mathrm{UV}$} & \colhead{$F_\mathrm{Ly\alpha}$} & \colhead{EW$_\mathrm{Ly\alpha}$} \\
    \colhead{} & \colhead{} & \colhead{} & \colhead{} & \colhead{} & \colhead{$\times10^{-19}$ erg s$^{-1}$ cm$^{-2}$} & \colhead{\AA} \\
    \colhead{(1)} & \colhead{(2)} & \colhead{(3)} & \colhead{(4)} & \colhead{(5)} & \colhead{(6)} & \colhead{(7)}
    }
    \startdata
    \hline
    \multicolumn{7}{c}{LAEs} \\ 
    \hline
    GS-9422 & 53.121755 & -27.797634 & 5.936 & $-19.83_{-0.03}^{+0.02}$ & $127.5^{+2.5}_{-2.1}$ & $235.1_{-4.2}^{+4.4}$\\
    GS-10056849 & 53.113511 & -27.772836 & 5.814 & $-18.73^{+0.07}_{-0.07}$ & $62.9^{+2.4}_{-2.2}$ & $178.6^{+5.4}_{-6.0}$ \\
    GS-113056 & 53.110517 & -27.798489 & 5.986 & $-17.41_{-0.18}^{+0.19}$ & $10.7^{+1.1}_{-1.1}$ & $136.7_{-12.9}^{+16.0}$\\
    GS-9365 & 53.162801 & -27.760838 & 5.921 & $-19.63_{-0.03}^{+0.03}$ & $152.4^{+12.5}_{-11.8}$ & $123.1_{-10.0}^{+8.3}$\\
    GS-138765 & 53.135802 & -27.765912 & 5.761 & $-16.65_{-0.19}^{+0.25}$ & $7.6^{+1.5}_{-1.4}$ & $113.2_{-20.1}^{+26.2}$\\
    GS-19606 & 53.176550 & -27.771108 & 5.889 & $-18.67_{-0.06}^{+0.06}$ & $76.8^{+3.7}_{-4.1}$ & $100.2_{-5.4}^{+4.8}$\\
    GS-40396 & 53.179863 & -27.808277 & 5.834 & $-18.67_{-0.07}^{+0.08}$ & $29.2^{+14.0}_{-6.4}$ & $84.5_{-17.3}^{+39.3}$\\
    GS-210003 & 53.131841 & -27.773774 & 5.779 & $-18.75_{-0.10}^{+0.11}$ & $42.9^{+1.4}_{-1.5}$ & $76.9_{-2.6}^{+2.8}$\\
    GS-40000170 & 53.136002 & -27.798489 & 5.777 & $-19.52_{-0.04}^{+0.04}$ & $56.8^{+11.1}_{-11.2}$ & $70.0_{-14.1}^{+12.6}$\\
    GS-9697 & 53.130437 & -27.802361 & 5.932 & $-19.43_{-0.07}^{+0.05}$ & $39.4^{+14.1}_{-11.7}$ & $70.0_{-20.8}^{+26.1}$\\
    GS-19342 & 53.160620 & -27.771613 & 5.974 & $-19.00_{-0.07}^{+0.07}$ & $33.2^{+2.4}_{-2.2}$ & $66.1_{-4.6}^{+4.8}$\\
    GS-108606 & 53.154201 & -27.805515 & 5.928 & $-16.28_{-0.38}^{+0.44}$ & $5.64^{+2.3}_{-1.8}$ & $58.7_{-19.6}^{+25.1}$\\
    GS-109389 & 53.122104 & -27.804292 & 5.789 & $-18.34_{-0.08}^{+0.10}$ &  $17.4^{+2.6}_{-2.3}$ & $49.7_{-6.5}^{+6.6}$\\
    GS-6002 & 53.110411 & -27.808923 & 5.937 & $-18.94_{-0.10}^{+0.10}$ & $31.1^{+1.7}_{-1.7}$ & $47.4_{-2.6}^{+2.4}$\\
    GS-201127 & 53.166853 & -27.804125 & 5.831 & $-19.48_{-0.07}^{+0.08}$ & $16.8^{+2.5}_{-2.4}$ & $36.0_{-5.8}^{+5.9}$\\
    GS-42905 & 53.140774 & -27.802177 & 5.916 & $-19.59_{-0.03}^{+0.03}$ & $36.2^{+6.8}_{-6.7}$ & $31.6_{-5.2}^{+5.2}$\\
    \hline
    \multicolumn{7}{c}{non-LAEs} \\
    \hline 
    GS-10005113 & 53.167303 & -27.802870 & 5.821 & $-18.05_{-0.12}^{+0.11}$ & & \\
    GS-10013618 & 53.119112 & -27.760802 & 5.944 & $-19.71_{-0.10}^{+0.09}$ & & \\
    GS-10013620 & 53.122590 & -27.760569 & 5.918 & $-19.81_{-0.06}^{+0.07}$ & & \\
    GS-10013704 & 53.126535 & -27.818092 & 5.920 & $-18.70_{-0.09}^{+0.09}$ & & \\
    GS-131737 & 53.199042 & -27.772580 & 5.889 & $-19.00_{-0.04}^{+0.05}$ & & \\
    GS-13466 & 53.191850 & -27.824993 & 5.780 & $-20.47_{-0.04}^{+0.04}$ & & \\
    GS-13638 & 53.141967 & -27.755230 & 5.829 & $-20.00_{-0.03}^{+0.03}$ & & \\
    GS-13639 & 53.125544 & -27.755050 & 5.792 & $-18.12_{-0.23}^{+0.29}$ & & \\
    GS-13651 & 53.149865 & -27.752831 & 5.920 & $-19.66_{-0.04}^{+0.05}$ & & \\
    GS-14133 & 53.172640 & -27.767058 & 5.986 & $-18.91_{-0.05}^{+0.06}$ & & \\
    GS-16431 & 53.105473 & -27.761147 & 5.922 & $-20.16_{-0.08}^{+0.07}$ & & \\
    GS-17213 & 53.161921 & -27.739928 & 5.927 & $-19.64_{-0.04}^{+0.05}$ & & \\
    GS-200277 & 53.109691 & -27.807613 & 5.937 & $-17.10_{-0.25}^{+0.28}$ & & \\
    GS-22251 & 53.154070 & -27.766072 & 5.798 & $-19.25_{-0.12}^{+0.11}$ & & \\
    GS-30080593 & 53.148846 & -27.809696 & 5.960 & $-17.50_{-0.09}^{+0.11}$ & & \\
    GS-3968 & 53.145047 & -27.816431 & 5.768 & $-18.30_{-0.12}^{+0.10}$ & & \\
    GS-40649 & 53.183929 & -27.807589 & 5.866 & $-19.20_{-0.03}^{+0.03}$ & & \\
    GS-41002 & 53.142273 & -27.806837 & 5.989 & $-18.69_{-0.07}^{+0.07}$ & & \\
    GS-4404 & 53.115372 & -27.814771 & 5.764 & $-19.45_{-0.08}^{+0.07}$ & & \\
    GS-51778 & 53.138363 & -27.778780 & 5.935 & $-19.26_{-0.06}^{+0.07}$ & & \\
    GS-52721 & 53.202984 & -27.776131 & 5.887 & $-20.26_{-0.03}^{+0.03}$ & & \\
    GS-57895 & 53.145039 & -27.762823 & 5.933 & $-19.73_{-0.04}^{+0.04}$ & & \\
    GS-9301 & 53.166728 & -27.804240 & 5.828 & $-21.79_{-0.01}^{+0.01}$ & & \\
    GS-9457 & 53.173240 & -27.795674 & 6.002 & $-20.17_{-0.03}^{+0.03}$ & & \\
    GS-9696 & 53.177520 & -27.802520 & 5.864 & $-19.25_{-0.05}^{+0.07}$ & & \\
    GS-9735 & 53.199380 & -27.796274 & 5.985 & $-20.03_{-0.04}^{+0.04}$ & & \\
    GS-9833 & 53.142082 & -27.779848 & 5.921 & $-20.44_{-0.03}^{+0.03}$ & & \\
    GS-99671 & 53.126643 & -27.817731 & 5.923 & $-18.24_{-0.11}^{+0.10}$ & & \\
    \enddata
    \tablecomments{(1) JADES NIRSpec ID; (2) right ascension; (3) declination; (4) redshift; (5) rest-frame UV magnitude; (6) \lya\ flux; (7) \lya\ equivalent width.}
\end{deluxetable*}

\end{appendix}

\bibliography{main}{}
\bibliographystyle{aasjournalv7}



\end{document}